\documentclass[aip,jcp,reprint,amsmath,amssymb,showpacs,superscriptaddress]{revtex4-2}

\usepackage{appendix}
\usepackage{amsmath}
\usepackage{amsfonts}
\usepackage{amssymb}
\usepackage{xcolor}
\usepackage{verbatim}
\usepackage{graphicx}
\usepackage{physics}
\usepackage{mathtools}
\usepackage{soul}
\usepackage{algorithm}
\usepackage{algpseudocode}
\usepackage{array}
\usepackage{booktabs}
\definecolor{darkblue}{rgb}{0,0,0.6}
\usepackage[colorlinks,linkcolor=darkblue,citecolor=darkblue,urlcolor=darkblue]{hyperref}
\usepackage{tikz,standalone}
\usetikzlibrary{calc}
\usetikzlibrary{shapes.geometric}
\usetikzlibrary{matrix,positioning}

\newcommand{\vct}[1]{\boldsymbol{\mathit{#1}}}
\newcommand{\beq}{\begin{equation}}
\newcommand{\eeq}{\end{equation}}

\begin{document}

\title{Policy-guided Monte Carlo on general state spaces: Application to glass-forming mixtures}
\author{Leonardo Galliano}
\affiliation{Dipartimento di Fisica, Universit\`a di Trieste, Strada Costiera 11, 34151, Trieste, Italy}

\author{Riccardo Rende}
\affiliation{International School for Advanced Studies (SISSA), Via Bonomea 265, 34136, Trieste, Italy}
\author{Daniele Coslovich}
\email{dcoslovich@units.it}
\thanks{This article may be downloaded for personal use only. Any other use requires prior permission of the author and AIP Publishing. This article appeared in the Journal of Chemical Physics and may be found at \url{https://doi.org/10.1063/5.0221221}.}
\affiliation{Dipartimento di Fisica, Universit\`a di Trieste, Strada Costiera 11, 34151, Trieste, Italy}

\date{\today}

\begin{abstract}

  Policy-guided Monte Carlo is an adaptive method to simulate classical interacting systems. It adjusts the proposal distribution of the Metropolis-Hastings algorithm to maximize the sampling efficiency, using a formalism inspired by reinforcement learning. In this work, we first extend the policy-guided method to deal with a general state space, comprising, for instance, both discrete and continuous degrees of freedom, and then apply it to a few paradigmatic models of glass-forming mixtures. We assess the efficiency of a set of physically inspired moves whose proposal distributions are optimized through on-policy learning. Compared to conventional Monte Carlo methods, the optimized proposals are two orders of magnitude faster for an additive soft sphere mixture but yield a much more limited speed-up for the well-studied Kob-Andersen model. We discuss the current limitations of the method and suggest possible ways to improve it.
  
\end{abstract}

\maketitle

\section{Introduction}

Since the seminal work of Metropolis \textit{et al.}~\cite{metropolisEquationStateCalculations1953}, Monte Carlo (MC) methods have provided a powerful tool to sample the equilibrium Boltzmann distribution of classical interacting systems.
In general, however, the actual number of iterations needed to reach an adequate sampling depends strongly on the system and on its thermodynamic conditions.
A broad range of methods has been developed to accelerate MC simulations, ranging from configurational bias moves for the simulation of long polymer chains~\cite{batoulis_kremer_1988, Grassberger_1997} to cluster moves involving several degrees of freedom for systems close to continuous phase transitions, both in lattice models~\cite{swendsenNonuniversalCriticalDynamics1987b, wangAntiferromagneticPottsModels1989, wolffCollectiveMonteCarlo1989} and on the continuum~\cite{dressClusterAlgorithmHard1995a, liuRejectionFreeGeometricCluster2004, makReverseMonteCarlo2007}.
Enhanced sampling techniques are thus part of the modern toolbox of computer simulations in statistical physics~\cite{frenkel_smit, landau_binder} and constitute a research topic on their own.

Recently, MC simulations have integrated ideas and concepts from machine learning, which provides a promising framework to boost enhanced sampling methods.
An interesting application involves optimizing the parameters of MC moves using automatic differentiation and stochastic gradient methods~\cite{Hastie_Tibshirani_Friedman_2016}.
This idea has been explored in different yet largely analogous ways, resulting in a new generation of adaptive MC methods aimed at maximizing the algorithm's performance~\cite{bojesenPolicyguidedMonteCarlo2018, zhao2019Ice,gabrie2022Adaptive, christiansen2023SelfTuning,  asghar2024efficient}.
However, despite the success of these and other related machine learning methods~\cite{wu2019Autoregressive, noe2019Boltzmann}, 
even in the proximity of phase transitions~\cite{marchand2022wavelet}, it is acknowledged that some of these methods struggle with systems exhibiting slow, glassy dynamics~\cite{ciarella2023Hard, jung2024Normalizing}.

Simulation of glassy systems provides indeed a hard benchcase for enhanced sampling techniques~\cite{berthierModernComputationalStudies2022}.
In the following, we will focus on simple models of glass-forming liquids, in which particles interact through a pairwise potential and a suitable mismatch between particle sizes frustrates the nucleation of the stable crystalline phase. 
Glass-forming liquids are characterized by a two-step, non-exponential relaxation and by a dramatic super-Arrhenius increase in structural relaxation times, which leads to a dynamic arrest at low temperatures~\cite{Berthier_Biroli_2011}.
Conventional MC simulations with local particle displacements then suffer from the same slowdown observed with molecular dynamics~\cite{berthierMonteCarloDynamics2007}.
Equilibration of glass-forming mixtures can, however, be significantly enhanced by performing swap moves, in which the chemical species of randomly selected pairs of particles are exchanged~\cite{grigeraFastMonteCarlo2001a}.
In 2017, Ninarello \textit{et al.}~\cite{ninarelloModelsAlgorithmsNext2017} showed that swap MC achieves orders-of-magnitude speedups in suitably designed models of polydisperse particles, which remain stable against crystallization even at very low temperatures.
These advances have enabled equilibrium computational studies of glassy systems at unprecedented low-temperature conditions~\cite{berthierModernComputationalStudies2022}.

The effectiveness of swap MC remains mostly limited to the rather special class of size-polydisperse particles, which can be thought of as simple models of glassy colloidal suspensions.
In binary mixtures that mimic metallic alloys, instead, swap moves either rapidly crystallize the system or have very low acceptance~\cite{ninarelloModelsAlgorithmsNext2017}.
Introducing a ternary component or a small fraction of continuously polydisperse particles, whose interaction parameters interpolate between the actual chemical components of the mixture, has brought some remarkable speed-ups~\cite{parmarUltrastableMetallicGlasses2020}.
However, it would be desirable to develop MC algorithms that enhance the sampling of glassy mixtures directly, without modifying their Hamiltonian or chemical composition.

To this end, we consider an adaptive MC algorithm, named policy-guided Monte Carlo (PGMC)~\cite{bojesenPolicyguidedMonteCarlo2018}, which delivers efficient MC moves as part of an optimization process inspired by reinforcement learning~\cite{suttonReinforcementLearningIntroduction2018}.
We will first generalize the method to the case of a general state space, comprising both continuous and discrete degrees of freedom, highlighting the similarities and differences with related approaches (Sec.~\ref{sec:algorithm}).
Our key application will be the simulation of two models of glass-forming mixtures for which normal swap MC is inefficient (Sec.~\ref{sec:results}).
We will show that for one of these mixtures PGMC identifies a biased swap move that is about two orders of magnitude more efficient than normal swaps.
We will conclude by discussing the current limitations of the method and suggesting possible ways to construct more efficient, collective moves (Sec.~\ref{sec:conclusions}).

\section{The algorithm}
\label{sec:algorithm}
The PGMC method was introduced by Bojesen in Ref.~\onlinecite{bojesenPolicyguidedMonteCarlo2018} to simulate classical spin models with a discrete state space.
Our goal in this section is to generalize the algorithm to sample probability distributions defined on a general state space.
This is relevant for the simulation of glass-forming mixtures, whose state space consists of both continuous (positional) and discrete (compositional) degrees of freedom.
We will start by reviewing the MC method for a general state space (Sec.~\ref{sec:MH}), then formulate the PGMC method building on the analogy with reinforcement learning (Sec.~\ref{sec:pgmc_theory_practice}), and finally discuss similarities and differences with other adaptive MC methods (Sec.~\ref{sec:related}).

\subsection{Metropolis-Hastings Monte Carlo}
\label{sec:MH}
To deal with general state spaces, we formulate the MC method in the setting of measure-theoretic probability~\cite{green1995Reversible, Billingsley_1995, tierney1998Metropolis, robert2004MonteCarlo}. 
Readers unfamiliar with this formalism may mentally replace the notation ``$P\left(\mathrm{d}x\right)$" with the more straightforward ``$p(x)\,\mathrm{d}x$".
Throughout this section and in Sec.~\ref{sec:pgmc_theory_practice}, we will strive to highlight the correspondence with the MC method for conventional probability distributions.

We consider a probability space $(\mathcal{X}, \Sigma, P)$. 
The set $\mathcal{X}$ denotes the sample space, \textit{i.e.}, the set of all possible outcomes, while $\Sigma$ is a collection of all possible events, which are subsets of $\mathcal{X}$.
$P$ is a probability distribution that assigns to any event $X\in\Sigma$ the probability of observing $X$.
For physical systems in thermal equilibrium, $\mathcal{X}$ corresponds to the phase space of all possible configurations. 
An event $X \in \Sigma$ can be thought of as a specific region within $\mathcal{X}$, such as a collection of selected configurations. 
In the present context, $P$ is the Boltzmann distribution, where $P\left(X\right)$ gives the probability of finding the system within the region $X$.

Monte Carlo methods allow to sample a probability distribution $P$ by generating a Markov chain of states $\left\{x_\mu\right\}_{\mu=1}^{M}$ distributed according to $P$, that is, $x_\mu\sim P$. 
The expected value of any observable $\mathcal A$ can then be estimated as the empirical average,
\beq
\int_{\mathcal X}\mathcal A(x)\,P\left(\mathrm{d}x\right)=\lim_{M\to+\infty}\frac{1}{M}\sum_{\mu=1}^{M}\mathcal A\left(x_\mu\right).
\label{eq:MonteCarloAverage}
\eeq
In the case of discrete degrees of freedom, the left-hand side of Eq.~\eqref{eq:MonteCarloAverage} can be written as 
\beq
\int_{\mathcal X}\mathcal A(x)\,P\left(\mathrm{d}x\right)=\sum_{x\in\mathcal X}\mathcal A(x)\,p\left(x\right),
\eeq
where $p(x)$ is the probability mass function in state $x$. 
On the other hand, when $P$ has a well-defined probability density function $p$, as is the case for continuous degrees of freedom, then
\beq
\int_{\mathcal X}\mathcal A(x)\,P\left(\mathrm{d}x\right)=\int_{\mathcal X}\mathcal A(x)p(x)\,\mathrm{d}x.
\eeq

In order to build the Markov chain, one needs to specify a transition kernel $K\left(x,X'\right)$, which quantifies the conditional probability of transitioning from state $x$ to a subset $X'\in\Sigma$ of the state space.
A sufficient condition for sampling the correct target distribution $P$ is that $K$ satisfies the detailed balance condition
\beq
P\left(\mathrm{d}x\right)K\left(x,\mathrm{d}x'\right)= P\left(\mathrm{d}x'\right)K\left(x',\mathrm{d}x\right).
\label{eq:DetailedBalance}
\eeq
Note that when both $P$ and $K$ have a well-defined density function, we have $P\left(\mathrm{d}x\right)=p(x)\,\mathrm dx$ and $K\left(x,\mathrm{d}x'\right)=k\left(x,x'\right)\,\mathrm dx'$, and the detailed balance relation~\eqref{eq:DetailedBalance} takes the more familiar form
\beq
p\left(x\right)k\left(x,x'\right)=p\left(x'\right)\,k\left(x',x\right).
\eeq

Among MC methods, the Metropolis-Hastings (MH) algorithm~\cite{metropolisEquationStateCalculations1953, zbMATH03349185} stands out for its effectiveness and widespread application.
In this scheme, a new state $x'$ is drawn from the proposal distribution $Q$. 
To satisfy detailed balance, the transition is accepted with probability 
\beq
\alpha\left(x,x'\right)=\min\left\{1,\,\frac{Q\left(x',\mathrm{d}x\right)}{P\left(\mathrm{d}x\right)}\frac{P\left(\mathrm{d}x'\right)}{Q\left(x, \mathrm{d}x'\right)}\right\},
\label{eq:MetropolisHastings}
\eeq
where $Q\left(x',\mathrm{d}x\right)/P\left(\mathrm{d}x\right)$ and $P\left(\mathrm{d}x'\right)/Q\left(x, \mathrm{d}x'\right)$ denote the Radon-Nikodym derivative~\cite{Billingsley_1995} of $Q$ with respect to $P$ and \textit{viceversa}.
When $P$ and $Q$ have well-defined probability mass or density functions $p$ and $q$, these derivatives are simply the ratios $q\left(x',x\right)/p\left(x\right)$ and $p\left(x\right)/q\left(x,x'\right)$, respectively~\footnote{We note that this holds even when $q$ and $p$ are ``generalized'' densities, as long as they are defined with respect to the \emph{same} measure. We also point out that, in some applications, the Radon-Nikodym derivative cannot be expressed as a ratio of $q$ and $p$. This occurs, for instance, in spatial point processes where the support of the proposal distribution changes dimension at each step~\cite{geyer1994Simulation,green1995Reversible}.}.

Within this proposal-rejection scheme, the transition kernel of the MH algorithm can be written as
\begin{multline}
    K\left(x,X'\right)=\int_{X'}\alpha\left(x,x'\right)\,Q\left(x,\mathrm dx'\right)+\\
    +\int_{\mathcal X}\big(1-\alpha\left(x,x'\right)\big)\,Q\left(x,\mathrm dx'\right)\delta\left(x,X'\right),
    \label{eq:MHGKernel}
\end{multline}
where $\delta$ is the identity kernel. 
The second term accounts for the case in which $x$ is already in $X'$ and the proposed transition is rejected.

\subsection{Policy-guided Monte Carlo: theory and practice}
\label{sec:pgmc_theory_practice}
Provided that the proposal distribution $Q$ guarantees ergodicity, the condition of detailed balance ensures that the MH algorithm eventually samples the desired distribution $P$. 
This leaves great freedom in the choice of the specific form of the proposal distribution. The general goal of adaptive MC methods, including PGMC, is to find an optimal proposal distribution that maximizes some measure of the sampling efficiency of the Markov chain.

PGMC builds upon an analogy with the standard problem of reinforcement learning~\cite{suttonReinforcementLearningIntroduction2018}, in which an \textit{agent} performs an \textit{action} $a: \mathcal X \to \mathcal X$, constantly changing the state of the system from $x$ to $x'$.
The agent is guided by a \textit{policy}, which specifies the probability of choosing a specific action given the state $x$ of the system.
In PGMC, the proposal of a new state is thus seen as an action and the proposal distribution $Q$ corresponds to the policy.
The acceptance term, Eq.~\eqref{eq:MetropolisHastings}, then plays the role of the dynamical rule determining the next state once the action has
been chosen.
Typically, in reinforcement learning, the policy and the dynamic rule are denoted as $\pi\left(a|x\right)$ and $p\left(x'|a,x\right)$, respectively.
In this work, we identify an action $a$ with a MC move, aligning our notation more closely with that of traditional MC simulations.
Finally, to quantify the performance of a policy, PGMC assigns a \textit{reward} $r\left(x,x'\right)$ to each transition $x\to x'$.
A natural choice is to set the reward to a measure of inverse correlation between states $x$ and $x'$, but other choices depending on the type of move are also possible.
Our objective function is the expected reward
\beq
J =\int_{\mathcal X^2}r\left(x,x'\right)\,K\left(x,\mathrm{d}x'\right)\,P\left(\mathrm{d}x\right),
\label{eq:ExpectedRewardKernel}
\eeq
where $\mathcal X^2=\mathcal X\times\mathcal X$.
In this work, we set $r\left(x,x\right)=0$ so that the system must change its state in order to accumulate some reward.
Using Eq.~\eqref{eq:MHGKernel}, we can then write the expected reward as an average over the policy,
\beq
J=\int_{\mathcal X^2}r\left(x,x'\right)\alpha\left(x,x'\right)\,Q\left(x,\mathrm{d}x'\right)\,P\left(\mathrm{d}x\right).
\label{eq:ExpectedReward}
\eeq

Our goal is to find a policy $Q^\star$ that maximizes the expected reward $J$.
To practically tackle the problem, we restrict the policy search to a family of distributions $Q_{\theta}$ parameterized by a real vector $\theta$.
Then, starting from an initial guess, we update $\theta$ iteratively according to the stochastic gradient ascent, procedure~\cite{suttonReinforcementLearningIntroduction2018}
\beq
\theta\leftarrow\theta + \eta\,\widehat{\grad J},
\label{eq:sga}
\eeq
where $\eta$ is the learning rate and $\widehat{\grad J}$ is a stochastic estimate of the actual gradient of the expected reward $J$ with respect to the parameters $\theta$.
In Appendix~\ref{subsec:OptimisationMethods}, we compare the efficiency of simple stochastic gradient ascent with alternative update rules, well-known in the context of reinforcement learning.

In order to use Eq.~\eqref{eq:sga}, we need to estimate the gradient of the expected reward.
Note that while $J$ is defined as an average over $Q$ and $P$ and, thus, can be easily estimated using Eq.~\eqref{eq:MonteCarloAverage}, estimating $\grad J$ is not straightforward.
In Appendix~\ref{subsec:estimating_grad}, we address this issue and derive an expression for $\grad J$ that can be sampled in the same way as the one in Eq.~\eqref{eq:ExpectedReward}, see Eqs.~\eqref{eq:MeasureTheoreticGradient} and~\eqref{eq:GradLogAlpha}.

The practical implementation of PGMC is very similar to that of the standard Metropolis-Hastings algorithm: 
in a conventional MC simulation, when designing a move, one must specify a method for sampling new states, given a \emph{fixed} policy $Q_{\theta}$.
The system then evolves by drawing successive states from $Q_{\theta}$ and accepting them with probability $\alpha$ given by Eq.~\eqref{eq:MetropolisHastings}.
The only difference that sets PGMC apart from MH is the integration of an optimization step at regular time intervals.
In practice, at each optimization step, a set of $M_x$ configurations is drawn from the standard MH procedure.
Then, for each of these configurations, a set of $M_{x'}$ new configurations is proposed according to the current policy $Q_{\theta}$. 
These samples can then be used to estimate the gradient $\grad J$ of the expected reward in $\theta$ as
\begin{widetext}
\beq
  \widehat{\grad J}=\frac{1}{M_x\,M_{x'}}\sum_{\mu=1}^{M_x}\sum_{\nu=1}^{M_{x'}}r\left(x_\mu,x'_\nu\right)\,\alpha_{\theta}\left(x_\mu,x'_\nu\right)\big(\grad\log q_{\theta}\left(x_\mu,x'_\nu\right)+\grad\log\alpha_{\theta}\left(x_\mu,x'_\nu\right)\big),
\label{eq:estimate_gradient}
\eeq
\end{widetext}
where $x_\mu\sim P(\cdot)$ and $x'_\nu\sim Q_{\theta}(x_\mu,\cdot)$, while $q_\theta\left(x,x'\right)$ is the probability density of $Q_{\theta}$ with respect to some reference measure and $\grad\log\alpha_\theta$ is given by Eq.~\eqref{eq:GradLogAlpha}.
The estimate $\widehat{\grad J}$ can then be used to update the policy according to Eq.~\eqref{eq:sga}.
Once convergence is achieved, the simulation proceeds with the policy $Q^\star$, which is expected to exhibit improved efficiency compared to the initial policy $Q_{\theta}$. 
We will provide concrete examples of ``generalized'' densities $q$ when discussing MC moves specifically designed for glass-forming mixtures.

Algorithm~\ref{algo:PGMC} provides the pseudo-code to implement the PGMC method for a general state space.
For simplicity, in this example, we perform an optimization step at each MC step of the standard MH algorithm.

\begin{algorithm}[H]
    \begin{algorithmic}[1]
        \State \textbf{Require:} target distribution $P$, policy $Q_{\theta}$
        \State \textbf{Inputs:} initial state $x_0$, initial parameters $\theta$, number of MC steps $n_{step}$, number of $x$ samples per optimization step $M_x$, number of $x'$ samples per optimization step $M_{x'}$, learning rate $\eta$
        \For{$t\in 0,\dots,n_{steps}-1$}
            \For{$\mu\in1,\dots,M_x$}
                \State sample $x_\mu\sim K_{\theta}\left(x_t,\cdot\right)$
                \For{$\nu\in1,\dots,M_{x'}$}
                    \State sample $x'_{\nu}\sim Q_{\theta}\left(x_{\mu},\cdot\right)$
                \EndFor
            \EndFor
            \State estimate $\widehat{\grad J}$ from Eq.~\eqref{eq:estimate_gradient} using samples $\left\{x_\mu,x'_\nu\right\}$
            \State update $\theta\leftarrow\theta +\eta\widehat{\grad J}$
            \State sample $x_{t+1}\sim K_{\theta}\left(x_t,\cdot\right)$
        \EndFor
    \end{algorithmic}
    \caption{Policy-guided Monte Carlo}
    \label{algo:PGMC}
\end{algorithm}

Overall, the simulation framework of PGMC provides a flexible approach to automatically optimizing the parameters of MC moves. 
In contrast, conventional MC simulations require these parameters to be adjusted manually through a sequence of time-consuming preliminary simulations, see Sec.~\ref{sec:displacements} for an example.
We also anticipate that PGMC bears a strong structural similarity with other adaptive MC methods, which we briefly review in Sec.~\ref{sec:related}.

\subsection{Comparison with adaptive Monte Carlo methods}
\label{sec:related}
The idea of optimizing proposal distributions in MC simulations has been explored several times in the past~\cite{roberts1997WeakCA, haario1999AdaptivePD, latuszynski2011Adaptive}.
For instance, in Ref.~\onlinecite{andrieu2008Tutorial}, Andrieu and Thoms reviewed a class of algorithms called adaptive Monte Carlo methods, in which simple transition probabilities are adjusted to maximize different measures of performance. 
In this context, Pasarica and Gelman~\cite{pasarica2010AdaptiveScaling} introduced an objective function analogous to the expected reward defined in Eq.~\eqref{eq:ExpectedReward}. 
Namely, they chose a reward function $r\left(x,x'\right)$ that measures the square distance between consecutive states $x$ and $x'$ in the Markov chain.
These adaptive Monte Carlo methods update transition probabilities by employing optimization techniques that rely on simple grid maximization or analytical derivatives of the proposal density~\cite{pasarica2010AdaptiveScaling}.

Recent progress in machine learning has popularized the use of automatic differentiation~\cite{rall1981automatic}, which enables efficient evaluation of derivatives for generic functions.
Thanks to these advances, it is possible to formulate general update rules applicable to any well-behaved proposal distribution. 
As we have seen in the previous sections, the PGMC method builds precisely on these ideas and the analogy with reinforcement learning.
More recently, Christiansen \textit{et al}.~\cite{christiansen2023SelfTuning} applied a similar concept to fine tune the parameters of the so-called Hamiltonian Monte Carlo~\cite{duane1987HMC}, which considers short molecular dynamics trajectories as deterministic proposals in the MH algorithm~\footnote{This method is analogous to hybrid Monte Carlo, well-known in the context of the simulation of liquids.}.
In this framework, the timestep and the number of integration steps are adjusted to optimize yet another measure of distance between consecutive states using back-propagation.

In his original formulation of the PGMC method, Bojesen introduced the notion of ``chain policies'', in which multiple moves are combined into a single one~\cite{bojesenPolicyguidedMonteCarlo2018}.
However, determining the proposal density for these composite moves is not trivial, due to the many pathways through which the system could transition between two states when multiple random moves are combined.
In principle, this issue can be resolved by imposing the stronger condition of path-wise detailed balance~\cite{nilmeier2011Superdetailed,tamagnone2024CoarseGrained}.
However, it remains unclear to us how to define chain policies that simultaneously act on different degrees of freedom, \textit{e.g.}, positions and spins.
For this reason, while the original PGMC algorithm~\cite{bojesenPolicyguidedMonteCarlo2018} relied on short trajectories to estimate gradients, our version of PGMC employs single transitions $x\to x'$ and thus resembles more closely the ``reinforce'' algorithm for continuing tasks~\cite{suttonReinforcementLearningIntroduction2018}.
The main difference with the standard version of this latter algorithm lies in the fact that the acceptance term in Eq.~\eqref{eq:ExpectedReward} is policy-dependent, thereby introducing an additional term in the gradient of the expected reward.
Note that, within our approach, it is still possible to design collective moves, but one must be able to explicitly write the probability of transitioning between two consecutive states, independently of the specific path. 
We will illustrate an example of such a move in Sec~\ref{sec:softswap}.

While all the above methods focus on optimizing proposal distributions with a small number of parameters, recent studies have opted for large neural network models to parameterize transition probabilities~\cite{zhao2019Ice,gabrie2022Adaptive,asghar2024efficient}. 
For example, Gabri\'e \textit{et al.}~\cite{gabrie2022Adaptive} employed a normalizing flow as a proposal distribution, which generates a new configuration $x'$ independently of the current state $x$.
In their work, they tune the normalizing flow to minimize the Kullback-Leibler divergence~\cite{kullback1951} with the target distribution. 
As noted by Andrieu and Moulines in Ref.~\onlinecite{andrieu2006Ergodic}, this approach is equivalent to maximizing the average acceptance probability and corresponds to setting $r\left(x,x'\right)=1-\delta\left(x-x'\right)$ in Eq.~\eqref{eq:ExpectedReward}.
We think that PGMC represents a comprehensive framework that integrates these ideas and can be employed to develop general transition distributions, using more complex parametrizations than the ones we will actually investigate in this work.

\section{Numerical experiments}
\label{sec:results}

We now use the PGMC method to simulate classical models of interacting particles that exhibit glassy dynamics.
The state of the system is specified by $x=\left(\vct{r}, \vct{s}\right)$, where $\vct r=\left\{\vec r_1,\dots,\vec r_N\right\}$ represents the vector containing the positions of each particle and $\vct s=\left\{s_1,\dots,s_N\right\}$ denotes the list of their chemical species.
The interaction between particles is described by a potential energy surface $U(\vct{r},\vct{s})$, which we will define in Sec.~\ref{sec:details}.
The models are assumed to be in thermal equilibrium with a heat bath at temperature $T$, hence the states are distributed according to the Boltzmann distribution $P$, with a fixed global chemical composition.

The PGMC method is very general and allows us to optmize any parametrized move.
The challenge lies in designing suitable rewards and policies for efficient sampling of configuration space.
In Secs~\ref{sec:displacements} and~\ref{sec:softswap}, we introduce a series of policies that are easy to sample and parametrize.
We will build on physical intuition to design policies that aim for more efficient sampling compared to simple displacement and swap moves~\cite{ninarelloModelsAlgorithmsNext2017}.
In Sec.~\ref{sec:performance}, we will provide an overall performance assessment and quantify the extent to which these policies can speed up the simulation of our benchmark models.

\subsection{Simulation details}
\label{sec:details}

We simulate systems composed of $N=216$ particles in a cubic box with periodic boundary conditions. 
We focus on three discrete mixture models: two different binary mixtures~\cite{bernuSoftsphereModelGlass1987, Kob_Andersen_1995} composed of $N_A$ particles of type $A$ and $N_B$ particles of type $B$, as well as a ternary mixture~\cite{parmarUltrastableMetallicGlasses2020} that includes $N_C$ particles of type $C$.
Particles interact via a pairwise potential $u_{\alpha\beta}\left(r\right)$, for $\alpha,\beta\in\left\{A, B,C\right\}$.
All these glass-forming mixtures are also fairly robust against crystallization, at least in the temperature range accessible to conventional MC simulations.

The soft sphere model~\cite{bernuSoftsphereModelGlass1987} is a 50:50 mixture with $N_A=N_B=108$. The pair potential is given by an inverse power law
\beq
u_{\alpha\beta}(r)=\epsilon\left(\frac{\sigma_{\alpha\beta}}{r}\right)^{12} .
\label{eq:ipl}
\eeq
The number density is $\rho=0.5342$, and we fix $\epsilon=1.0$, $\sigma_{AA}=1.0$, $\sigma_{AB}=\sigma_{BA}=1.2$ and $\sigma_{BB}=1.4$.
We found that the onset of glassy dynamics occurs around $T\approx 0.25$ for this model, see Sec.~\ref{sec:performance}.
Note that the size ratio of this mixture is larger than the one used in earlier simulation studies employing the swap MC method~\cite{grigeraFastMonteCarlo2001a, ninarelloModelsAlgorithmsNext2017}.

The binary Kob-Andersen mixture~\cite{Kob_Andersen_1995} is a prototypical glass-forming mixture.
We study systems composed of $N_A=173$ and $N_B=43$ particles at the usual density $\rho=1.2$.
The interaction is given by the Lennard-Jones potential,
\beq
u_{\alpha\beta}(r)=4\epsilon_{\alpha\beta}\left[\left(\frac{\sigma_{\alpha\beta}}{r}\right)^{12}-\left(\frac{\sigma_{\alpha\beta}}{r}\right)^{6}\right],
\label{eq:LennardJones}
\eeq
with $\epsilon_{AA}=1.0$, $\epsilon_{AB}=\epsilon_{BA}=1.5$, $\epsilon_{BB}=0.5$, $\sigma_{AA}=1.0$, $\sigma_{AB}=\sigma_{BA}=0.8$, and $\sigma_{BB}=0.88$.
The onset of glassy dynamics occurs around $T \approx 1.0$~\cite{sastrySignaturesDistinctDynamical1998}.

Finally, we also study a variant of the Kob-Andersen mixture~\cite{parmarUltrastableMetallicGlasses2020}, which is characterized by three chemical species with $N_A=144$ and $N_B=N_C=36$, at a density $\rho=1.2$.
The interaction parameters of the additional $C$ species interpolate between those of the $A$ and $B$ species of the original Kob-Andersen mixture.
Namely, particles interact again with the Lennard-Jones potential Eq.~\eqref{eq:LennardJones}, where the interaction parameters for particles $A$ and $B$ are the same as in the original Kob-Andersen model, while 
$\epsilon_{AC}=\epsilon_{CA}=1.25$, $\epsilon_{BC}=\epsilon_{CB}=1.0$, $\epsilon_{CC}=0.75$,
$\sigma_{AC}=\sigma_{CA}=0.9$, $\sigma_{BC}=\sigma_{CB}=0.84$, and $\sigma_{CC}=0.94$.
Thanks to this modification, the model can be simulated efficiently using the swap MC algorithm even below the so-called mode-coupling crossover temperature, which is around $0.34$~\cite{parmarUltrastableMetallicGlasses2020}.
However, swap MC simulations struggle to equilibrate the system below $T\approx 0.3$.

In each of the three models, we truncate and shift the potential at a cutoff distance $r^c_{\alpha\beta}=2.5\sigma_{\alpha\beta}$.

We perform simulations that combine several moves in the same run, such as the displacement of a single particle and the exchange of two particles of different species (swap).
Each move is independently associated with a reward $r$ and a policy $Q_{{\theta}}$, which corresponds to a specific parametrization of the proposal distribution.
We also assign to each move a weight, which quantifies the probability of performing that type of move at each MC step. 
For the sake of simplicity, we fix these weights at the beginning of the simulation and we optimize each policy independently.
Once a type of move has been chosen, a transition $x\to x'$ is proposed according to the policy $Q_{{\theta}}$ and it is accepted according to Eq.~\eqref{eq:MetropolisHastings}.

We perform an optimization step for each move at every MC sweep (corresponding to $N$ MC steps), which defines our time unit.
In order to estimate $\widehat{\grad J}$ at each optimization step, we employ the sampling strategy presented in Algorithm~\ref{algo:PGMC}: starting from configuration $x_t$, we draw $M_x$ samples from $K_\theta\left(x_t,\cdot\right)$, by performing MC sweeps.
For each of these samples, we draw $M_{x'}$ samples from $Q_{\theta}\left(x_\mu,\cdot\right)$.
This process yields a total of $M_xM_{x'}$ samples for estimating $\widehat{\grad J}$ using Eq.~\eqref{eq:estimate_gradient}.
The choice of $M_x$ and $M_{x'}$ depends on the temperature and on the specific optimization algorithm used to update the parameters (see Appendix~\ref{subsec:OptimisationMethods}).
This additional optimization procedure effectively increases the computational cost by approximately a factor of $M_{x'}$.
In practice, we tuned the values of $M_x$, $M_{x'}$ and the learning rate $\eta$ so that the parameters converged in roughly the same amount of time as a standard MC simulation for data production.
Once convergence is achieved, we perform standard MC simulations to compute the relevant structural and dynamical quantities, performing standard consistency checks, see Appendix~\ref{appendix:checks}.
The code for these simulations has been implemented in Julia, using the \texttt{Enzyme}~\cite{enzyme2020} package for automatic differentiation.

\subsection{Displacement moves}\label{sec:displacements}
To illustrate the PGMC method in the simplest possible case, we consider a straightforward displacement move, which corresponds to shifting a single particle $i$ from position $\vec r_i$ to $\vec r^{\,\prime}_i$.
Such a move is sampled from the policy $Q_{\sigma}$, where particle $i$ is chosen uniformly with probability $1/N$, and $\vec r^{\,\prime}_i$ is sampled from a Gaussian distribution with mean $\vec r_i$ and variance $\sigma^2$.
All other particles must remain fixed, and no particle can change its species.
For this move, the generalized density can be written as 
\beq
\label{eqn:displacement}
q_{\sigma}\left(x,x^\prime\right)=\begin{cases}
  \dfrac{\exp\left(-\frac{\lVert|\vec{r}^{\,\prime}_i\!- \vec{r}_i\rVert^2}{2\sigma^2}\right)}{N\!\left(2\pi\sigma^2\right)^{3/2}}&\exists! i\,|\,\vec{r}^{\,\prime}_i\!\neq\vec{r}_i\\
  0&\text{otherwise.}
\end{cases}
\eeq

Here, $\sigma$ is the only free parameter of the policy.
We set the reward to $r\left(x,x'\right)=\sum_{i}\norm{\vec r^{\,\prime}_i-\vec r_i}^2$, so that small displacements (which are always accepted but inefficient) are discouraged.
This is equivalent to minimizing the lag-one autocorrelation among configurations~\cite{pasarica2010AdaptiveScaling,christiansen2023SelfTuning}.

\begin{figure}
    \includegraphics[width=0.48\textwidth]{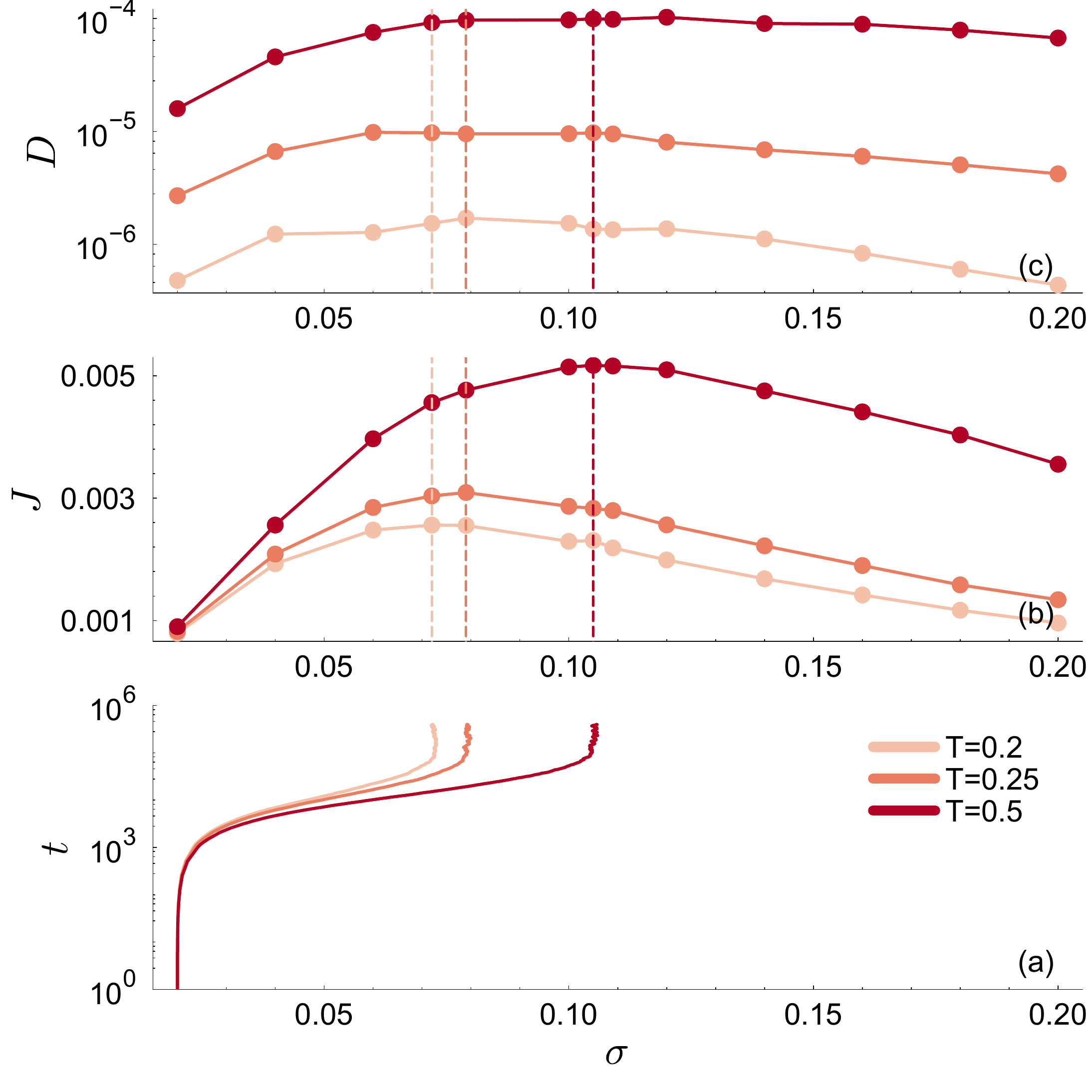}
    \caption{Parameter optimization for displacement moves:
    (a) time evolution of $\sigma$ during a PGMC simulation for three selected temperatures;
    (b) expected reward $J$ as a function of $\sigma$ obtained from standard MC simulations;
    (c) diffusion coefficient $D$ as a function of $\sigma$ obtained from standard MC simulations. 
    The dashed lines in panels (b) and (c) indicate the optimal parameters identified by PGMC in panel (a).}
    \label{fig:displacements}
\end{figure}

Figure~\ref{fig:displacements} illustrates how the algorithm converges to the optimal value of $\sigma$ for the soft sphere mixture.
In particular, panel (a) shows the evolution of $\sigma$ during optimizations at different temperatures, ranging from the normal liquid ($T=5$) to the moderately glassy regime ($T=0.2$).
The converged value $\sigma^*$ corresponds to a maximum of the expected reward $J$, see panel (b).
We note, however, that minimizing short-time correlations does not ensure a faster exploration of configuration space at long times.
To quantify the latter, we compute instead the mean square displacement
\beq
\Delta^2(t)=\frac{1}{N}\sum_{i=1}^{N}\expval{\norm{\vec r_i(t)-\vec r_i(0)}^2},
\eeq
and the corresponding diffusion coefficient 
\beq
D=\lim_{t\to\infty}{\frac{\Delta^2(t)}{6t}}.
\eeq
As a reference, we obtained $D$ from several independent MC simulations using fixed values of $\sigma$.
In Fig.~\ref{fig:displacements}(c), we see that the maximum of the expected reward is indeed very close to the maximum of the diffusion coefficient $D(\sigma)$.
This shows, quite strikingly, that optimizing the efficiency of the individual MC step also maximizes the long time diffusion, possibly thanks to a coupling between short- and long-time dynamics.

\subsection{Biased displacements}\label{sec:biased_displacements}
\begin{figure}
    \includegraphics[width=0.48\textwidth]{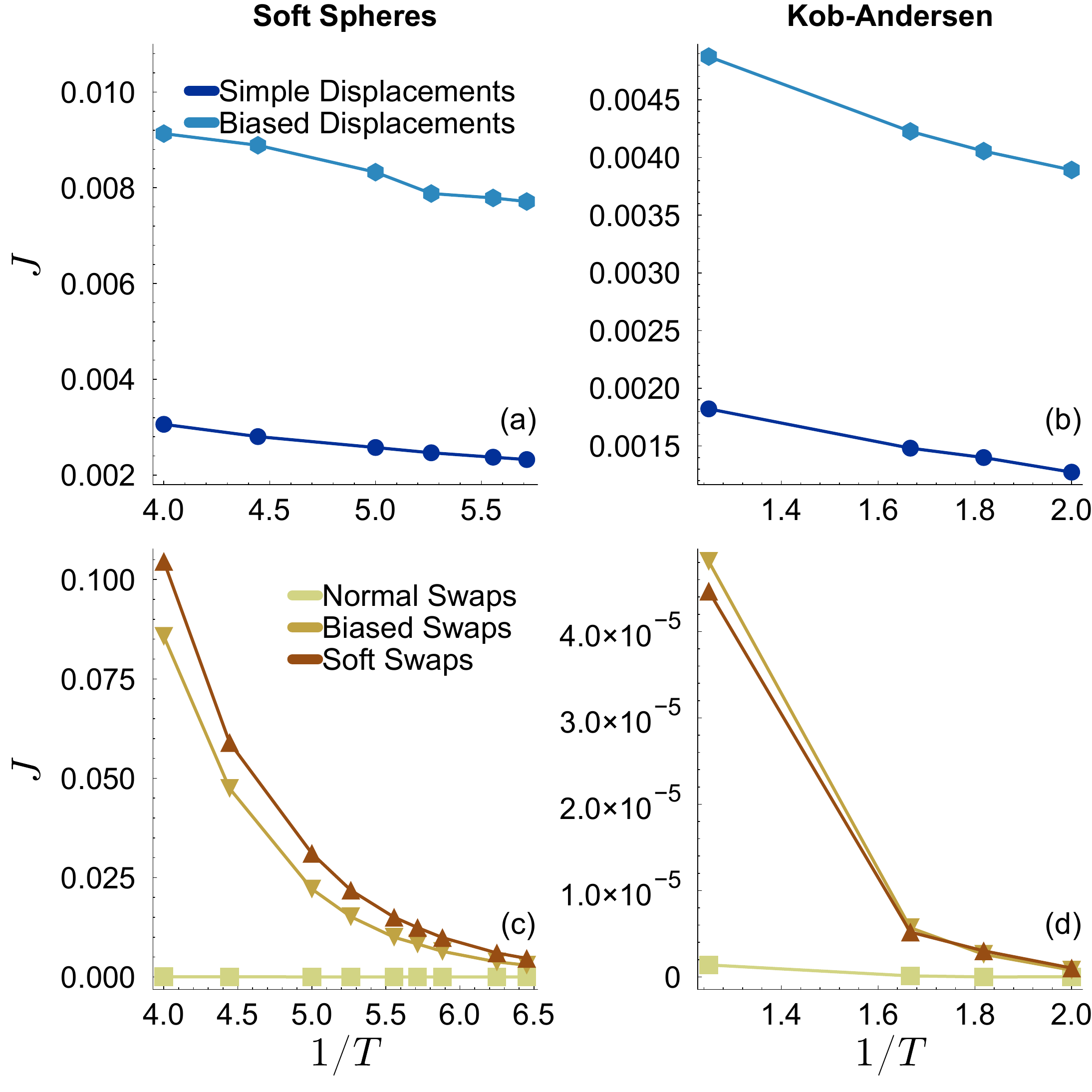}
    \caption{Average rewards as a function of $1/T$ for displacement moves (panels (a) and (b)) and swap moves (panels (c) and (d)).
      The left and right panels display results for the soft sphere and Kob-Andersen models, respectively.
    }
    \label{fig:rewards}
\end{figure}

Displacement moves can be improved by encoding information on the local environment around a particle directly into the policy.
This is a generalization of the so-called biased moves, well-known in standard MC methods~\cite{frenkel_smit}.
As a natural extension of Eq.~\eqref{eqn:displacement}, we thus consider the following biased displacement: the proposed position $\vec r^{\,\prime}_i$ is drawn from a Gaussian whose mean
includes a bias term aligned with the local force $\vec f_i(x)$ acting on particle $i$:
\begin{multline}
q_{\sigma,\lambda}\left(x,x^\prime\right)=\\
=\begin{cases}
  \dfrac{\exp\left(-\frac{\lVert\vec{r}^{\,\prime}_i\!- \vec{r}_i-\lambda\sigma^2\vec f_i(x)/T\rVert^2}{2\sigma^2}\right)}{N\!\left(2\pi\sigma^2\right)^{3/2}}&\exists! i\,|\,\vec{r}^{\,\prime}_i\!\neq\vec{r}_i\\
  0&\text{otherwise.}
\end{cases}
\end{multline}
This is a reformulation in the PGMC framework of the ``smart Monte Carlo'' algorithm proposed by Rossky \textit{et al.}~\cite{rosskySmartMC}, the only difference being the inclusion of the parameter $\lambda$ to adjust the amount of bias.
Such a policy incorporates more detailed information about the current state $x$ of the system and leads to better performance compared to  simple displacements.
This is illustrated in the top panels of Fig.~\ref{fig:rewards} for both the soft sphere and Kob-Andersen models for temperatures below the onset of glassy dynamics.
The expected rewards of biased displacements are about a factor 2 larger than those of simple displacements, irrespective of temperature.
We will analyze the actual increase of performance in Sec.~\ref{sec:performance}, by analyzing the exploration of configuration space.

\subsection{Biased swaps}\label{sec:biased_swaps}
A swap move involves exchanging the species of randomly chosen pairs of different kinds of particles~\cite{frenkel_smit}.
Despite their remarkable effectiveness in equilibrating polydisperse systems~\cite{ninarelloModelsAlgorithmsNext2017}, swap moves are inadequate for most binary mixtures.
The crux is that binary mixtures with a size ratio close to unity can be simulated efficiently with swap MC~\cite{grigeraFastMonteCarlo2001a}, but also have a strong tendency to crystallize around the mode-coupling crossover temperature~\cite{ninarelloModelsAlgorithmsNext2017}.
By contrast, mixtures with a larger size ratio or non-additive interactions, such as the ones used in this work, are more robust against crystallization, but the energy cost of enlarging the smaller particle dramatically suppresses the acceptance of swap moves, see Appendix~\ref{appendix:energy}.

\begin{figure}
    \includegraphics[width=0.48\textwidth]{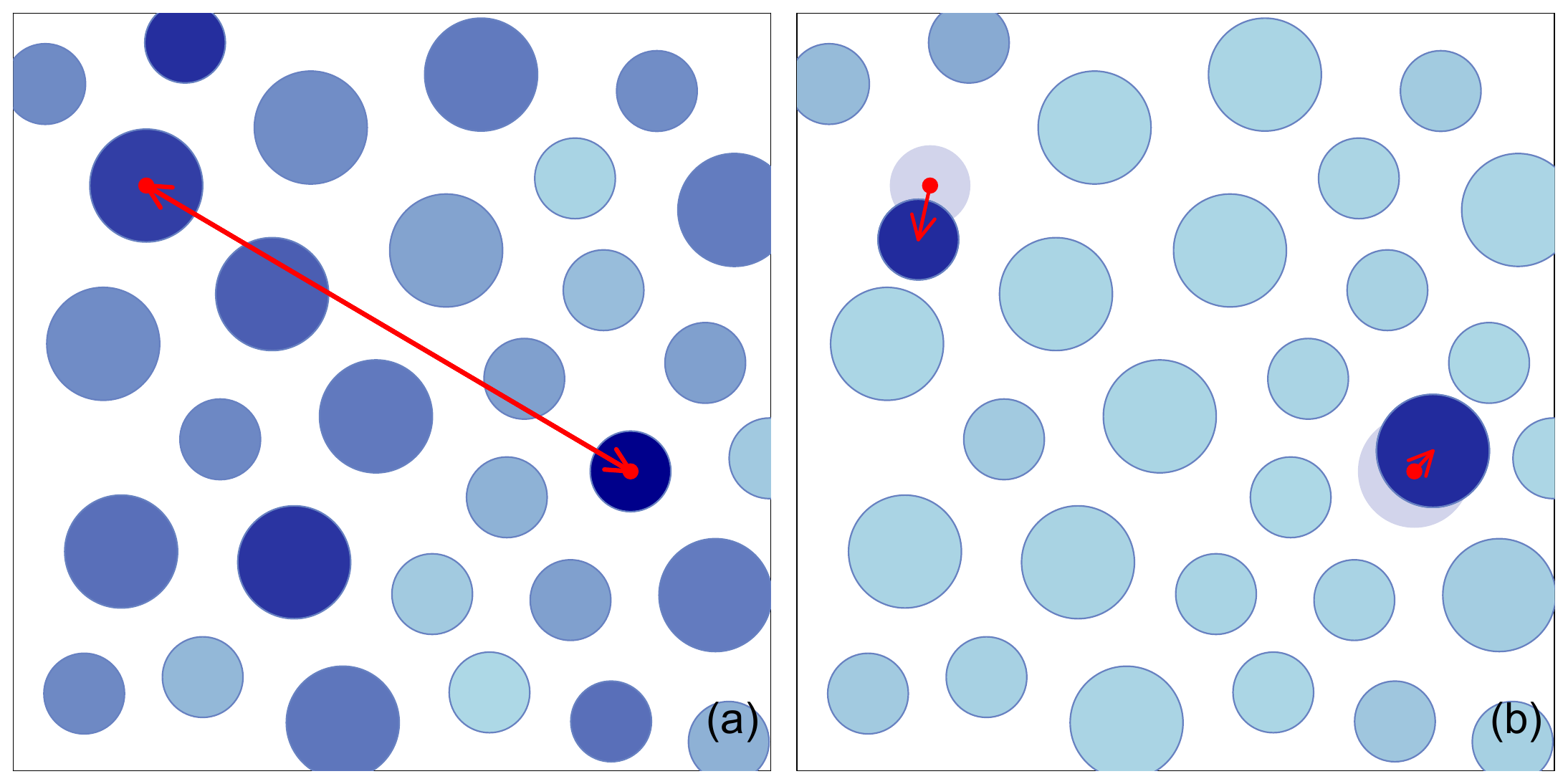}
    \caption{Sketches of (a) biased and (b) soft swap moves.
      In panel (a), the color code indicates the probability of selecting the particle, with darker blue indicating a higher probability; the arrow connects the particles whose swap is attempted.
      In panel (b), we show the additional biased displacements, indicated by the arrows, attempted after the swap.}
    \label{fig:softswap}
\end{figure}

In the standard implementation of the swap MC algorithm, the particle pairs are uniformly sampled from all the possible $N_A N_B / 2$ pairs.
One possible strategy to improve the method is to incorporate information about the local structure into the pair selection, identifying those pairs whose swap is more likely to be accepted.
A sketch of such a biased swap move is shown in Fig.~\ref{fig:softswap}(a).

\begin{figure}
    \includegraphics[width=0.48\textwidth]{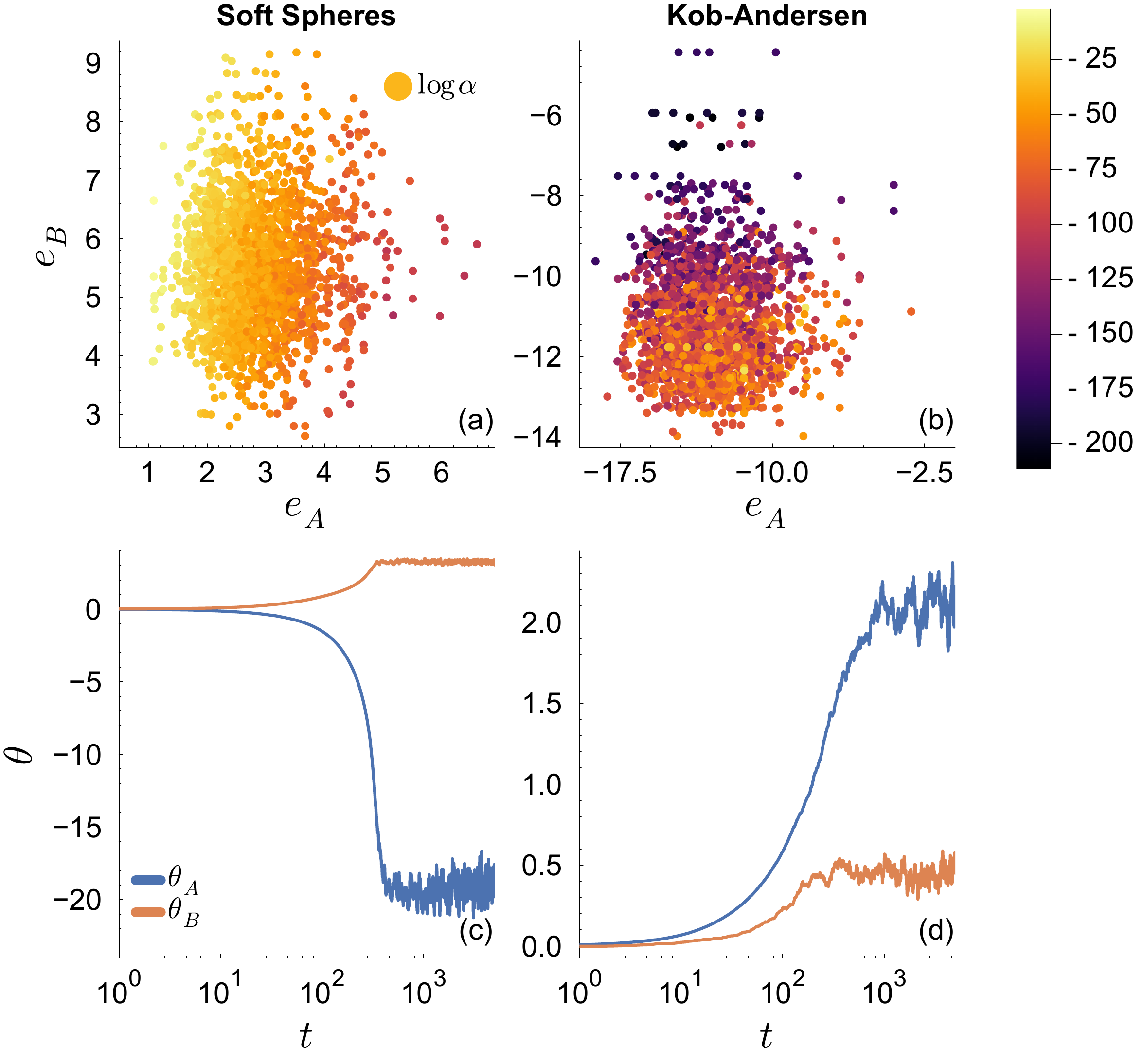}
    \caption{
      Upper panels: scatter plot of the local energies $\left(e_A, e_B\right)$ of pairs of particles before a normal swap from representative configurations at $T=0.3$ and $T=0.8$ in (a) the soft sphere and (b) the Kob-Andersen models, respectively.
      The points are color-coded by the acceptance $\log\alpha$ of the attempted normal swap.
      Lower panels: time evolution of $\theta_A$ and $\theta_B$ during a PGMC simulation for (c) soft sphere and (d) Kob-Andersen models.}
    \label{fig:swap}
\end{figure}

As a simple test of this idea, we check whether the local energy can be used to identify the most favorable pairs to swap.
Figure~\ref{fig:swap} shows a scatter plot of pairs of local potential energies $(e_i(x), e_j(x))$, with particles $i$ and $j$ of species $A$ and $B$, respectively.
For each model, we consider equilibrium samples from MC simulations close to the respective onset temperature.
We color-code each pair according to the acceptance of the swap in the absence of bias, \textit{i.e.}, $\log\alpha = \min\left\{0, \log p\left(x'\right)-\log p\left(x\right)\right\}$, where $y$ corresponds to the configuration in which $i$ and $j$ are swapped.
In the soft sphere mixture, we observe a systematic increase in acceptance as $e_A$ decreases: pairs with low local energy of the small particle have a higher probability of being swapped.
In the Kob-Andersen mixture, the typical acceptance is lower by several orders of magnitude and, crucially, it is not easy to identify the pairs with a larger $\log\alpha$ just by looking at their potential energies: the pairs with the highest probability of being swapped do not clearly lie at the boundaries of the energy distribution.
These results indicate that introducing a simple bias on the local energy could make a more effective swap move for the soft sphere, much less so for the Kob-Andersen mixture.

The PGMC method provides a natural framework for designing a move that exploits this information. 
We thus introduce a biased swap, in which particles $i\in A$ and $j\in B$ are chosen from a categorical distribution with a bias linked to their respective local energies.
Given the sensitivity of the acceptance to local energy, a natural parametrization for the proposal density of such a move is
\begin{widetext}
\beq
  q_{\theta_A,\theta_B}\left(x,x^\prime\right)=\begin{cases}
    \dfrac{\exp\left({\theta_Ae_i(x)}\right)}{\sum_{k\in A}\exp\left({\theta_Ae_k(x)}\right)} \dfrac{\exp\left({\theta_Be_j(x)}\right)}{\sum_{k\in B}\exp\left({\theta_Be_k(x)}\right)}&\exists!i\in A\,|\,s'_i\neq s_i\wedge\exists!j\in B\,|\,s'_j\neq s_j\\
    \,0&\text{otherwise.}
  \end{cases}
  \label{eq:biased_swaps}
\eeq
\end{widetext}
The two parameters $\theta_A$ and $\theta_B$ control the sign and strength of the bias.
To promote those swaps that are more likely to be accepted, we maximize the acceptance of the move and set the reward to $r\left(x,x'\right)=1-\delta\left(x-x'\right)$.
Note that the computational complexity of the move is $\mathcal O(N)$, unless the swap is restricted to a finite subset of (possibly neighboring) pairs.
This is due to the fact that sampling a categorical distribution with $N_c$ categories takes $\mathcal O(\log N_c)$ with binary search, but evaluating the normalization factor in Eq.\eqref{eq:biased_swaps} requires $\mathcal O(N_c)$ operations.
In principle, this additional overhead could be mitigated by implementing a cache to store the normalization factor. 
However, with each policy update, such a cache would have to be completely recalculated. 

The results of the optimization of the parameters $\theta_A$ and $\theta_B$ are shown in Fig.~\ref{fig:swap}(c) and (d).
At least for the soft sphere model, the values of optimal parameters confirm our physical intuition: the gradient of $\log\alpha$ in Fig.~\ref{fig:swap}(a) indicates that it is favorable to swap pairs in which the $A$-particle has low energy and indeed we find $\theta_A<0$ and $\abs{\theta_A}>\abs{\theta_B}$.
Note that it would be inefficient for the algorithm to just make $\theta_A$ as small as possible: the optimal value results from the balance between the weights of the forward and inverse moves within Eq.\eqref{eq:MetropolisHastings}.
The optimized parameters for the Kob-Andersen mixture are instead both positive, which probably reflects the difficulty of finding an effective bias for this model.
As we shall see below, in fact, biased swap moves are inefficient for this model, see Appendix~\ref{appendix:energy} for additional details.
More sophisticated approaches may be required to enhance swap moves in the Kob-Andersen mixture.

\subsection{Soft swaps}\label{sec:softswap}
The primary source of energy cost in a swap move is due to the enlargement of the smaller particle, see Appendix~\ref{appendix:energy}.
A potential strategy to mitigate this issue is to lower the local energy of such a particle, thereby reducing the energy difference between the two states.
We thus introduce a composite MC move, where the two swapped particles undergo a biased displacement right after the swap.
These displacements occur prior to the acceptance evaluation and potentially allow the system to tunnel between energy barriers.

To sample such a composite move, particles $i\in A$ and $j\in B$ are again chosen from a categorical distribution, as in the biased swap move. 
Then, the two particles are independently shifted to new positions $\vec r_i^{\,\prime}$ and $\vec r_j^{\,\prime}$ as in the biased displacement move.
A sketch of the additional step involved in such as a soft swap move is shown in Fig.~\ref{fig:softswap}(b).

The expected rewards $J$ obtained for soft swaps are shown in Figs.~\ref{fig:rewards}(c) and (d).
Compared to biased swaps, soft swaps provide no improvement for the Kob-Andersen mixture.
We observe instead a marginal increase in $J$ for the soft sphere mixture with respect to biased swaps.
Such a difference may be deemed negligible when simulating glassy liquids, where large speed-ups are needed in order to achieve equilibration at low temperatures~\cite{ninarelloModelsAlgorithmsNext2017}.
Nevertheless, this result shows that the PGMC framework can be used to implement complex moves in a constructive way, building on physical intuition as well.

\subsection{Performance assessment}
\label{sec:performance}

\begin{figure}[!htb]
    \includegraphics[width=0.48\textwidth]{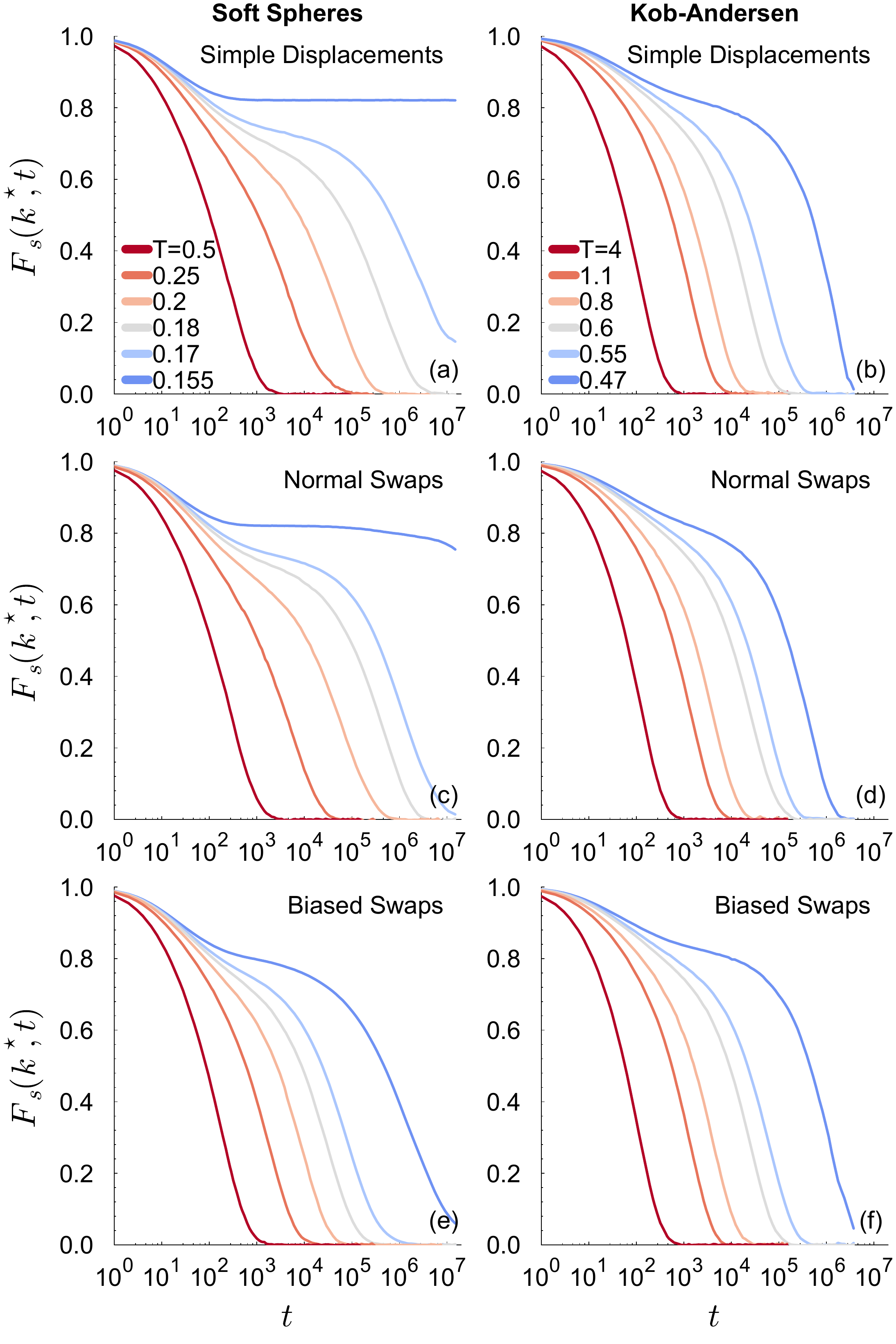}
    \caption{
      Comparison of self intermediate scattering functions $F_s(k^*,t)$ for selected temperatures in the soft sphere (left panels) and Kob-Andersen models (right panels).
      Panels (a) and (b): simple displacements. Panels (c) and (d): normal swaps.  Panels (e) and (d): biased swaps. 
    }
    \label{fig:fskt}
\end{figure}

To compare the performance of the policies described in the previous sections, we consider standard time-dependent correlation functions used to quantify structural relaxation in glass-forming liquids~\cite{berthierMonteCarloDynamics2007}.
In particular, we compute the self intermediate scattering function
\beq
F_s\left(k,t\right)=\frac{1}{N}\sum_{i=1}^{N}\expval{\exp\left(\mathrm{i}\vec k\cdot\left(\vec r_i(t)-\vec r_i(0)\right)\right)}
\eeq
at the first peak $k=k^*$ of the total structure factor
\beq
S\left(k\right)=\frac{1}{N}\sum_{i,j}\expval{\exp\left(\mathrm{i}\vec k\cdot\left(\vec r_i-\vec r_j\right)\right)}.
\eeq
We define the structural relaxation time $\tau_\alpha$ as $F_s\left(k^*,\tau_{\alpha}\right)=1/\mathrm{e}$.
This definition provides a well-defined measure of how efficient the exploration of configuration space is, even in the presence of particle swaps~\cite{ninarelloModelsAlgorithmsNext2017}.

In Fig.~\ref{fig:fskt}, we present the $F_s(k^*,t)$ obtained from MC simulations of the soft sphere and Kob-Andersen mixture using simple displacements, simple displacements combined with swaps, and simple displacements combined with biased swaps.
When combining displacements and swap moves, we set the probability of selecting a swap among the possible moves to 0.1. 
Each simulation is performed using the optimal parameters obtained in a preliminary PGMC simulation. 
From the figure we recognize the typical features of glassy dynamics: below the onset temperature $T_o$, the correlation functions develop a plateau due to the cage effect of the neighboring particles.
The results for the soft sphere mixture illustrate well the effectiveness of the PGMC framework.
At the lowest temperature $T=0.155$, neither simple displacements nor normal swaps can equilibrate this model.
This can be appreciated by the fact that the plateau in $F_s(k^*, t)$ stretches beyond the observation time scale of our simulations.
However, simulations with biased swaps (and soft swaps, not shown) allow the correlation function to relax below $1/\mathrm e$ even at this temperature. 

Biased and soft swaps are, by contrast, of little use for the Kob-Andersen mixture.
As shown in the right panels of Fig.~\ref{fig:fskt}, the decay of $F_s(k^*, t)$ occurs on similar time scales irrespective of the simulation method.
In the Kob-Andersen mixture, in fact, swap moves have negligible acceptance probabilities and neither biased nor soft swaps can promote their acceptance.
For this mixture, a small performance improvement can be achieved by introducing biased displacements, see below.

\begin{figure*}[!tb]
  \includegraphics[width=0.8\textwidth]{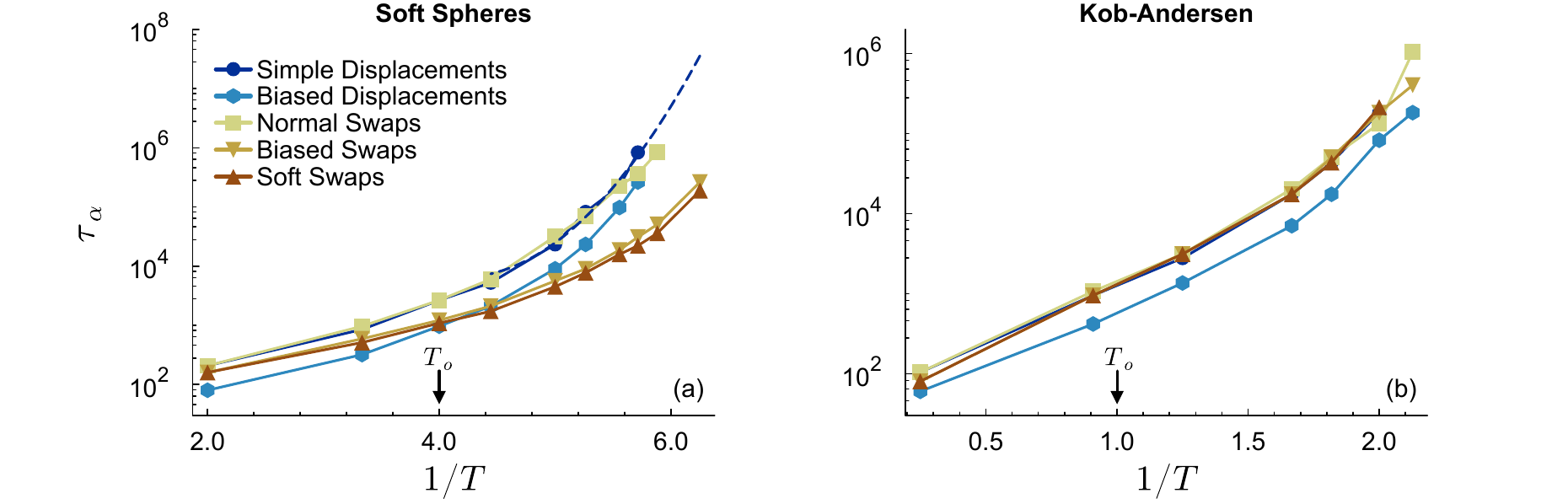}
  \caption{
    Structural relaxation time $\tau_\alpha$ for all policies in (a) soft sphere and (b) Kob-Andersen models.
    The dashed line in panel (a) indicates a fit to Eq.~\eqref{eqn:parabolic}.
    In panel (a), the symbols for simple and biased displacements lie almost on top of each other, as do those for biased and soft swaps. In panel (b), the symbols for all moves except bias displacements lie almost on top of each other.
  In both panels, the vertical arrows mark the onset temperature.
  }
  \label{fig:tau}
\end{figure*}

To provide a more quantitative evaluation of the performance, we show in Fig.~\ref{fig:tau} the temperature dependence of the structural relaxation times $\tau_\alpha$ for all the policies.
For the soft sphere model, the performance of biased and soft swaps is nearly identical.
Below the onset temperature $T_o$, these improved swap moves provide a substantial reduction in structural relaxation times compared to displacement moves and normal swaps.
At the temperatures at which all methods can achieve equilibrium, we estimate the speed-up of biased and soft swaps to be about two orders of magnitude.
To estimate the speed-up at lower temperatures, we extrapolate the $\tau_\alpha$ obtained from displacements and normal swaps using the parabolic law~\cite{Elmatad_Chandler_Garrahan_2009}
\begin{equation}
  \label{eqn:parabolic}
  \tau_\alpha = \tau_0 \exp[J^2\left(\frac{1}{T} - \frac{1}{T_o}\right)^2] \, ,
\end{equation}
fitted in the range of temperatures below the onset temperature.
At the lowest temperature at which biased and soft swaps can fully equilibrate the system, \textit{i.e.}, $T=0.16$, we estimate that the speed-up is still about 2 orders of magnitude.
While this result is less spectacular than the one achieved by normal swaps in polydisperse particles models close to the laboratory glass transition temperature~\cite{ninarelloModelsAlgorithmsNext2017, Berthier_Charbonneau_Ninarello_Ozawa_Yaida_2019}, it is nonetheless a significant achievement.
These configurations could be used, for instance, as starting points to study the equilibrium dynamics of this mixture below the mode-coupling crossover temperature~\cite{Coslovich_Ozawa_Kob_2018}.
The results for the Kob-Andersen mixture in panel (b) show instead that none of the swap moves brings any performance improvement over simple displacements.
For this model, biased displacements nonetheless provide a speed-up of about a factor of 2, irrespective of temperature.

\begin{figure}[!bp]
  \includegraphics[width=0.48\textwidth]{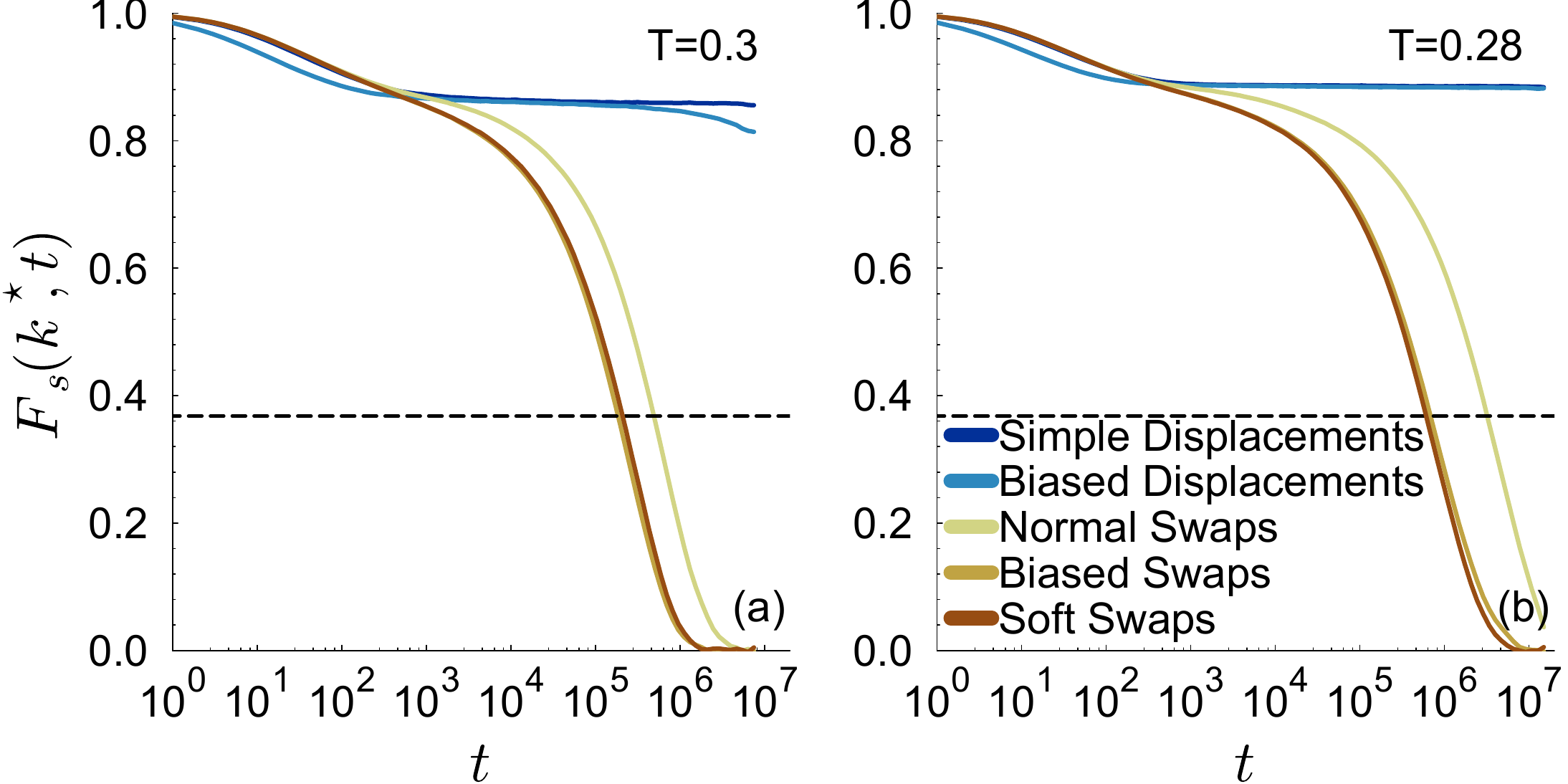}
  \caption{Self intermediate scattering functions $F_s(k^*,t)$ obtained from MC simulations for all policies in the ternary variant of the Kob-Andersen mixture at (a) $T=0.3$ and (b) $T=0.28$. The horizontal dashed line is drawn at $1/\mathrm{e}$. 
    In both panels, the lines for simple and biased displacements lie almost on top of each other, as do those for biased and soft swaps.
  }
  \label{fig:ternary}
\end{figure}

The physical interpretation of the above results is that biased swaps are effective in mixtures for which the acceptance of \textit{normal} swaps is non-negligible, at least close to the onset temperature.
In such systems, swap MC will eventually become inefficient upon cooling~\cite{ninarelloModelsAlgorithmsNext2017}, as the standard Metropolis acceptance rate drops with decreasing temperature.
Introducing an energy bias raises acceptance and shifts the numerical glass transition to a lower temperature.
To illustrate these ideas and corroborate our results, we consider a ternary variant of the Kob-Andersen mixture, which has been equilibrated successfully using normal swap MC even below the mode-coupling crossover temperature.
We simulate the model around the lowest temperature at which normal swap MC achieves equilibration, $T=0.28$.
In Fig.~\ref{fig:ternary}, we show the $F_s(k^*,t)$ obtained for this model with all the available policies.
We see that biased swaps reduce the relaxation times by almost an order of magnitude around this temperature.
Biased swaps are thus a potential candidate to push simulations of this model close to experimental time scales~\cite{berthierModernComputationalStudies2022}.

Finally, to take into account the different computational costs of the policies, we also evaluated the typical wall time necessary to carry out a single move.
Although the timing depends, in general, on implementation details and hardware, the \textit{ratios} between wall times of different moves are fairly robust and can be estimated from the number of single-particle energy evaluations.
We found that the overhead required by complex policies does not severely impact the performance of our PGMC code:
biased displacements take about 1.15 times as long as simple displacements, while biased swaps take approximately 1.4 times as long as normal swaps.
Soft swaps are more costly as they take about twice as long as biased swaps.
Focusing on the soft sphere model at $T=0.16$, we found that biased (soft) swaps achieve an effective speed-up in the range 90-100 (60-70), compared to normal swaps.

\section{Conclusions}
\label{sec:conclusions}

In this paper, we have formally extended the PGMC method~\cite{bojesenPolicyguidedMonteCarlo2018} to deal with a general state space, comprising for instance both continuous and discrete degrees of freedom.
The result is a flexible MC simulation framework that optimizes non-symmetric proposal distributions to maximize sampling efficiency.
The key motivation behind this extension was to simulate glass-forming mixtures, whose dynamics become sluggish at low temperatures, which represents a hard challenge for enhanced sampling techniques~\cite{berthierModernComputationalStudies2022, jung2024Normalizing}.
We have shown that a simple set of biased moves, based on elementary particle swaps and optimized through the PGMC algorithm, increase the sampling efficiency by two orders of magnitude in a mixture of additive soft spheres, for which simple displacement and swap moves become soon inadequate when lowering the temperature.
By contrast, the same moves achieved essentially no speed-up in the well-studied Kob-Andersen mixture.
In this model, we found a $2\times$ speed-up with biased displacements, which is, however, insufficient in the context of glass transition studies~\cite{berthierModernComputationalStudies2022}.

The origin of this system-dependence can be traced back to the different acceptance rates of elementary swaps.
The bias introduced in the PGMC moves raises efficiency by identifying those rare pairs of particles whose swap is more likely to be accepted.
In the soft sphere model and in another ternary mixture we studied, there exists a temperature range over which elementary swaps have non-negligible acceptance and biased swaps successfully shift the numerical glass transition to a lower temperature.
However, in the Kob-Andersen mixture the swap acceptance rate is simply too low at any temperature below the onset of slow dynamics.

Similar biased MC moves can be easily adapted to other models. For instance, one could design a biased swap move for polydisperse systems that discourages swaps when the particle radii are too close.
With regard to systems where local energies are not relevant, such as hard spheres, alternative local descriptors could be used~\cite{Tanaka_Tong_Shi_Russo_2019}.
In principle, one could consider complex parametrizations relating the moves' parameters to the local structure, \textit{e.g.}, through neural networks that learn a better local descriptor.
In this scenario, more sophisticated algorithms might be required to ensure reasonable convergence times.
To this end, it would be interesting to implement adaptive MC algorithms suitable for advanced reinforcement learning techniques, see Ref.~\onlinecite{wang2024RLMH} for very recent work in this direction.
Nevertheless, we think it is unlikely that these more complex parametrizations could significantly speed up simulations, as long as the algorithm is built on simple local moves.

A promising alternative approach involves implementing more complex collective moves in which multiple degrees of freedom are updated simultaneously. 
This idea has been recently explored by Berthier and coworkers~\cite{ghimenti2024irreversible}, who constructed a collective swap move using an event-chain MC approach~\cite{krauthEventchainMonteCarlo2021}.
PGMC could provide a tool to build such collective moves autonomously using more flexible policies with a large number of parameters. 
In this context, neural networks have proven to be a versatile tool for modeling probability distributions, as demonstrated in various physical applications~\cite{wu2019Autoregressive,noe2019Boltzmann,gabrie2022Adaptive,carleo2017Quantum,viteritti2023Transformer}. 
In particular, auto-regressive networks or normalizing flows could be advantageous for our case, providing arbitrarily complex policies that can be directly sampled -- a key requirement in MC simulations.
Implementing such policies in the PGMC framework represents an interesting extension to be explored in future work.

\section*{Acknowledgments} 

We thank T. A. Bojesen for the useful discussions about the formulation of chain policies and the implementation of the code.
We also extend our gratitude to L. Viteritti for the insightful conversations on the method and effective ways to present it.
The article was produced with co-funding from the European Union - Next Generation EU.

\section*{Author declarations}

\subsection*{Conflict of Interest}
The authors have no conflicts to disclose.
\subsection*{Author contributions}
\textbf{Leonardo Galliano:} Writing - review \& editing (equal).
\textbf{Riccardo Rende:} Writing - review \& editing (equal).
\textbf{Daniele Coslovich:} Writing - review \& editing (equal).

\section*{Data availability}
The data that support the findings of this study will be openly available after publication in Zenodo at \url{https://doi.org/10.5281/zenodo.11396665}.

\appendix

\section{Estimating $\grad J$}
\label{subsec:estimating_grad}
In this section, we devise a method to estimate $\grad J$, taking cues from the policy gradient theorem for continuing tasks~\cite{suttonReinforcementLearningIntroduction2018} and likelihood ratio approaches~\cite{lecuyerFormalGradient1990}. 
Since $Q_{\theta}\left(x,\mathrm{d}x'\right)$ depends on $\theta$, one cannot simply differentiate with respect to $\theta$ inside the integral of Eq.~\eqref{eq:ExpectedReward}. 
Nonetheless, as stochastic gradient ascent~\eqref{eq:sga} requires an estimate of $\grad J$ on the \emph{current} parameters value $\theta=\theta_0$,  we can rewrite Eq.~\eqref{eq:ExpectedReward} using importance sampling as
\beq
J(\theta)=\int_{\mathcal X^2}r\!\left(x,x'\right)\alpha_{\theta}\!\left(x,x'\right)\rho_{\theta}\!\left(x,x'\right)\,Q_{\theta_0}\!\left(x,\mathrm{d}x'\right)P\!\left(\mathrm{d}x\right),
\label{eq:importance_sampling}
\eeq
where $\rho_\theta\left(x,x'\right)=Q_\theta\left(x,\mathrm{d}x'\right)/Q_{\theta_0}\left(x,\mathrm{d}x'\right)$.
Note that importance sampling can be done using any measure $\mu$ that dominates $Q_\theta$, meaning that $\mu\left(X'\right)=0$ implies $Q_\theta\left(x,X'\right)$.
This condition is typically referred to as $Q_\theta$ being absolutely continuous with respect to $\mu$ and it is denoted as $Q_\theta\ll\mu$~\cite{Billingsley_1995}.
Since the average in Eq.~\eqref{eq:importance_sampling} does not depend on $\theta$ anymore, assuming suitable regularity conditions~\cite{lecuyerFormalGradient1990}, we can now differentiate inside the integral in Eq.~\eqref{eq:importance_sampling} as
\begin{multline}
    \grad J(\theta)=\int_{\mathcal X^2}r\left(x,x'\right)\big(\alpha_{{\theta}}\left(x,x'\right)\grad\rho_{\theta}\left(x,x'\right)+\\
    +\rho_{\theta}\left(x,x'\right)\grad\alpha_{\theta}\left(x,x'\right)\big)\,Q_{\theta_0}\left(x,\mathrm{d}x'\right)\,P\left(\mathrm{d}x\right).
\end{multline}
Then, using $\grad f_{\theta}=f_{\theta}\grad\log f_{\theta}$ for $f_\theta\neq 0$ and $\rho_\theta\left(x,x'\right)\,Q_{\theta_0}\left(x,\mathrm{d}x'\right)=Q_{\theta}\left(x,\mathrm{d}x'\right)$, we get
\begin{multline}
    \grad J(\theta) = \int_{\mathcal X^2}r\left(x,x'\right)\alpha_{\theta}\left(x,x'\right)\big(\grad\log \rho_{\theta}\left(x,x'\right)+\\
    +\grad\log\alpha_{\theta}\left(x,x'\right)\big)\,Q_{\theta}\left(x,\mathrm{d}x'\right)\,P(\mathrm{d}x),
    \label{eq:MeasureTheoreticGradient}
\end{multline}
where
\begin{multline}
    \grad\log\alpha_{\theta}\left(x,x'\right)=\Theta\left(1-\alpha_\theta\left(x,x'\right)\right)\times\\
    \times\big(\grad\log \rho_\theta\left(x',x\right)-\grad\log \rho_\theta\left(x,x'\right)\big)
    \label{eq:GradLogAlpha}
\end{multline}
and $\Theta$ is the Heaviside function. 
Now Eq.~\ref{eq:MeasureTheoreticGradient} is written as an average over $P$ and $Q_{\theta}$, which can be readily sampled.

Note that when $Q_\theta$ and $Q_{\theta_0}$ have a probability density with respect to a common reference measure, \textit{i.e.}, when $\rho_{\theta}\left(x,x'\right)$ can be written as a ratio $q_{\theta}\left(x,x'\right)/q_{\theta_0}\left(x,x'\right)$ for some function $q$, then 
\beq
\grad\log \rho_{\theta}\left(x,x'\right)=\grad\log q_{\theta}\left(x,x'\right).
\eeq
This condition holds for most implementations of the MH algorithm, including the application we describe in this work.
We note that this holds even when $q$ is a ``generalized'' density, \textit{e.g.} when the degrees of freedom are both discrete and continuous.

\section{Optimization methods}
\label{subsec:OptimisationMethods}
In reinforcement learning, the optimization algorithm described by the update rule Eq.~\eqref{eq:sga} is commonly referred to as the ``vanilla'' policy gradient (VPG).
The underlying idea of VPG consists in maximizing a linear approximation of the average reward $J$ centered around the current parameter values $\theta_0$, while simultaneously penalizing substantial shifts in parameter space.
Although this approach naturally encourages policy updates towards regions yielding higher rewards, it may not fully account for the complex geometry of the policy space and lead to sub-optimal convergence rates~\cite{amariNaturalGradient1998,kakadeNaturalPolicyGradient2001,schulmanTRPO2017}.

The natural policy gradient (NPG) algorithm addresses this limitation by incorporating the geometry of the policy space into the learning process. 
This is achieved by replacing the Euclidean metric $\left\|\theta-\theta_0\right\|^2$ with the Fisher metric $\left(\theta-\theta_0\right)^TF\left(\theta_0\right)\left(\theta-\theta_0\right)$, where
\begin{multline}
F\left(\theta\right)=\int_{\mathcal X^2}\grad\log\rho_{\theta}\left(x,x'\right)\grad\log\rho_{\theta}\left(x,x'\right)^T\times\\\times Q_{\theta}\left(x,\mathrm{d}x'\right)\,P\left(\mathrm{d}x\right),
\label{eq:FisherMatrix}
\end{multline}
is the Fisher information matrix.
Intuitively, this choice can be motivated by observing that the Fisher metric coincides with the average of the second-order expansion of the Kullback-Leibler divergence $\mathcal D_{KL}\left(Q_{\theta_0}||Q_{\theta}\right)$ between $Q_{{\theta}_0}$ and $Q_{{\theta}}$, often interpreted as a measure of the distance between probability distributions~\cite{kakadeNaturalPolicyGradient2001}. 
One can show that penalizing such a distance in policy space leads to the update rule
\beq
\theta\leftarrow\theta + \eta\,\widehat{F}^{-1}\left(\theta\right)\widehat{\grad J},
\label{eq:nga}
\eeq
where $\widehat F({\theta})$ is a stochastic estimate of $F({\theta})$~\cite{kakadeNaturalPolicyGradient2001}.
Since the gradients $\grad\log\rho_\theta$ are already required by VPG at each step, estimating the Fisher matrix can be done at a minimal computational cost, as long as the number of parameters is not excessively large.

To accelerate convergence even further, one may explicitly require that the Fisher metric remain below a prescribed threshold $\epsilon$, rather than simply penalizing large steps in policy space.
Solving the corresponding constrained optimization problem yields the update rule
\beq
\theta\leftarrow\theta + \eta({\theta})\,\widehat{F}^{-1}\left(\theta\right)\widehat{\grad J},
\label{eq:anga}
\eeq
where the adaptive learning rate
\beq
\eta(\theta)=\sqrt{\frac{2\epsilon}{\widehat{\grad J}(\theta)^T \widehat{F}^{-1}(\theta) \widehat{\grad J}(\theta)}}
\label{eq:AdaptiveLearningRate}
\eeq
allows for larger steps when the natural gradient is small~\cite{schulmanTRPO2017}.
We call this version of the algorithm adaptive natural policy gradient (ANPG).

\begin{figure}
    \includegraphics[width=0.48\textwidth]{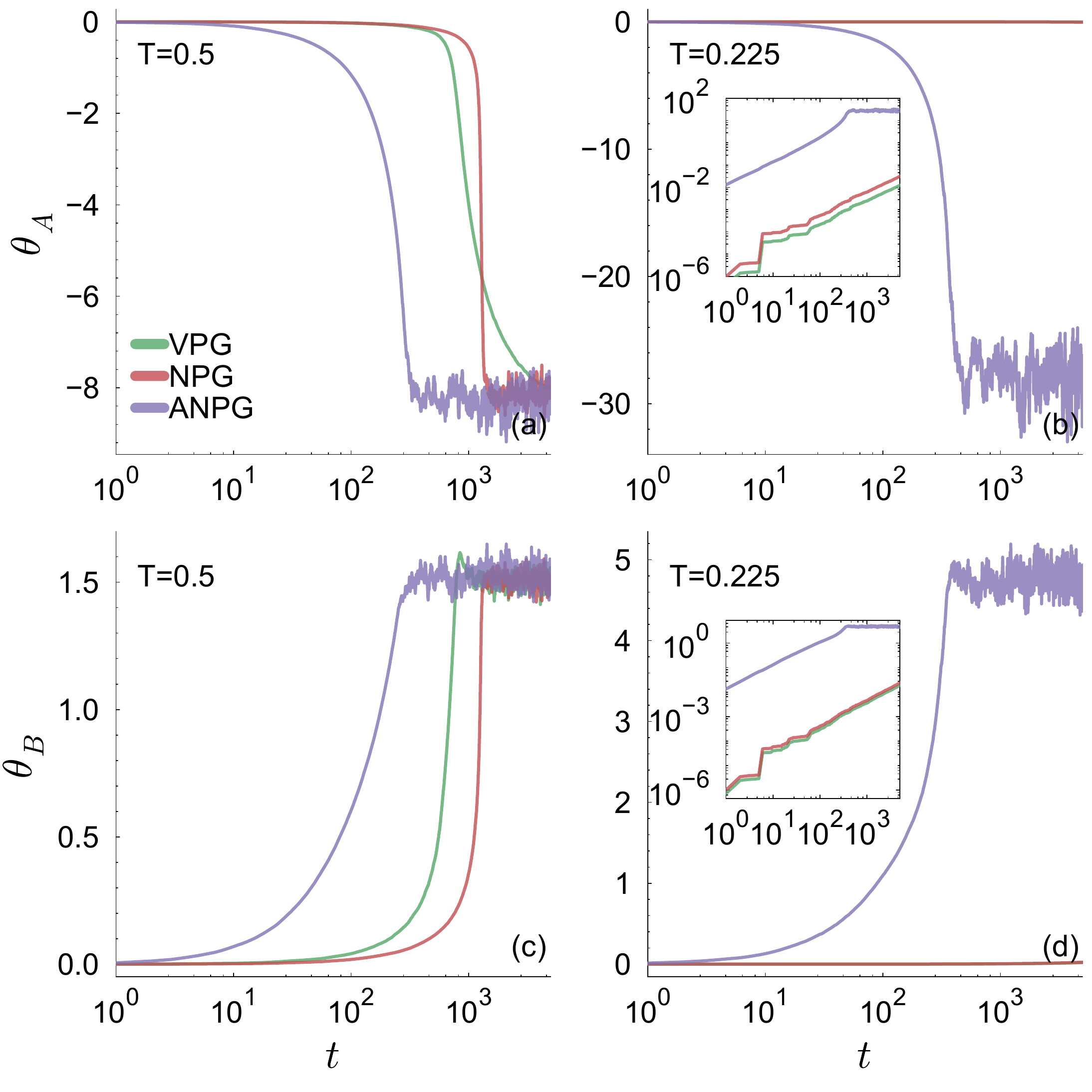}
    \caption{Overview of optimization methods at high temperature (left panels) and around the onset temperature (right panels) for the soft sphere model using biased swaps.
      Panels (a) and (b) show the time evolution of the parameter $\theta_A$, while panels (c) and (d) show the one of $\theta_B$.
    The insets in panels (b) and (d) display the same data on a log-log scale.}
    \label{fig:optimisation_methods}
\end{figure}

We compare the three methods for the soft sphere model in the normal liquid regime ($T=5$) and slightly below the onset temperature ($T=0.225$) in Fig.~\ref{fig:optimisation_methods}.
Since ANPG requires tuning a different parameter $\epsilon$ compared to the learning rate $\eta$ of the other two methods, we adjusted $\epsilon$ so that ANPG and NPG had similar parameter variance once convergence was achieved at $T=0.5$.
For VPG and NPG we set $\eta=0.1$, while for ANPG we used $\epsilon=10^{-4}$.
It is clear that ANPG leads to significantly faster convergence at low temperature compared to VPG and NPG.
Nevertheless, we point out that estimating $\widehat F(\theta)$ requires a higher number of samples compared to that needed for estimating $\widehat{\grad J(\theta)}$.
In this work, we used $M_x,M_{x'}\in\left[5,15\right]$ for VPG and $M_x,M_{x'}\in\left[20,50\right]$ for NPG and APG.

\section{Consistency checks}
\label{appendix:checks}

\begin{figure}[!tbp]
    \includegraphics[width=0.48\textwidth]{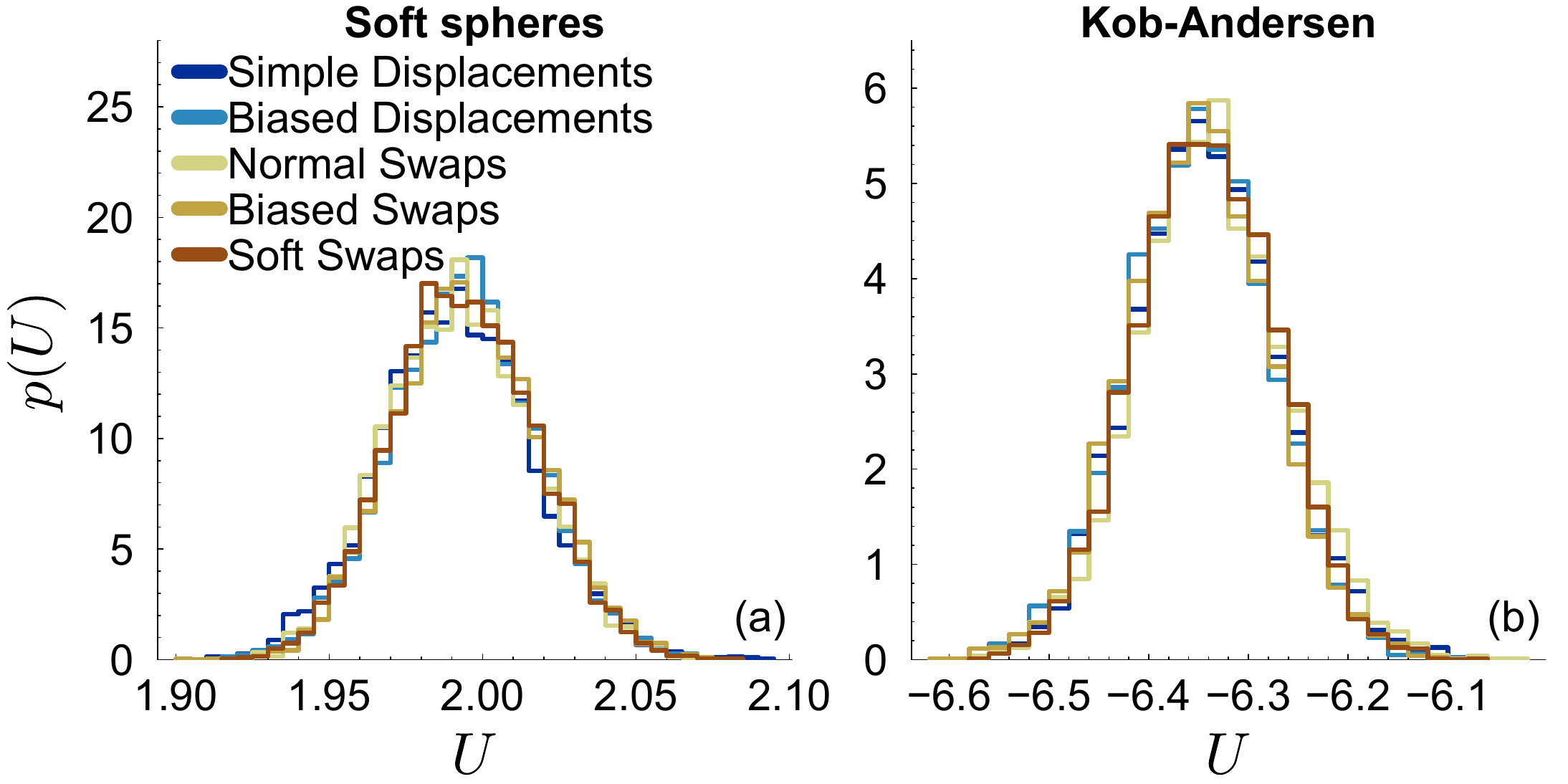}
    \caption{Potential energy distributions $p(U)$ obtained from MC simulations for all policies studied in (a) soft sphere and (b) Kob-Andersen models.}
    \label{fig:energy_distributions}
\end{figure}

As a consistency check, we verify that detailed balance is preserved  in our MC simulations by comparing in Fig.~\ref{fig:energy_distributions} the potential energy distributions for all policies.
Any discrepancies in these distributions would suggest an incorrect implementation of certain moves.
In our case, all policies yield overlapping energy distribution histograms in both models, indicating that each move samples the correct distribution.

\begin{figure}[!tbp]
    \includegraphics[width=0.48\textwidth]{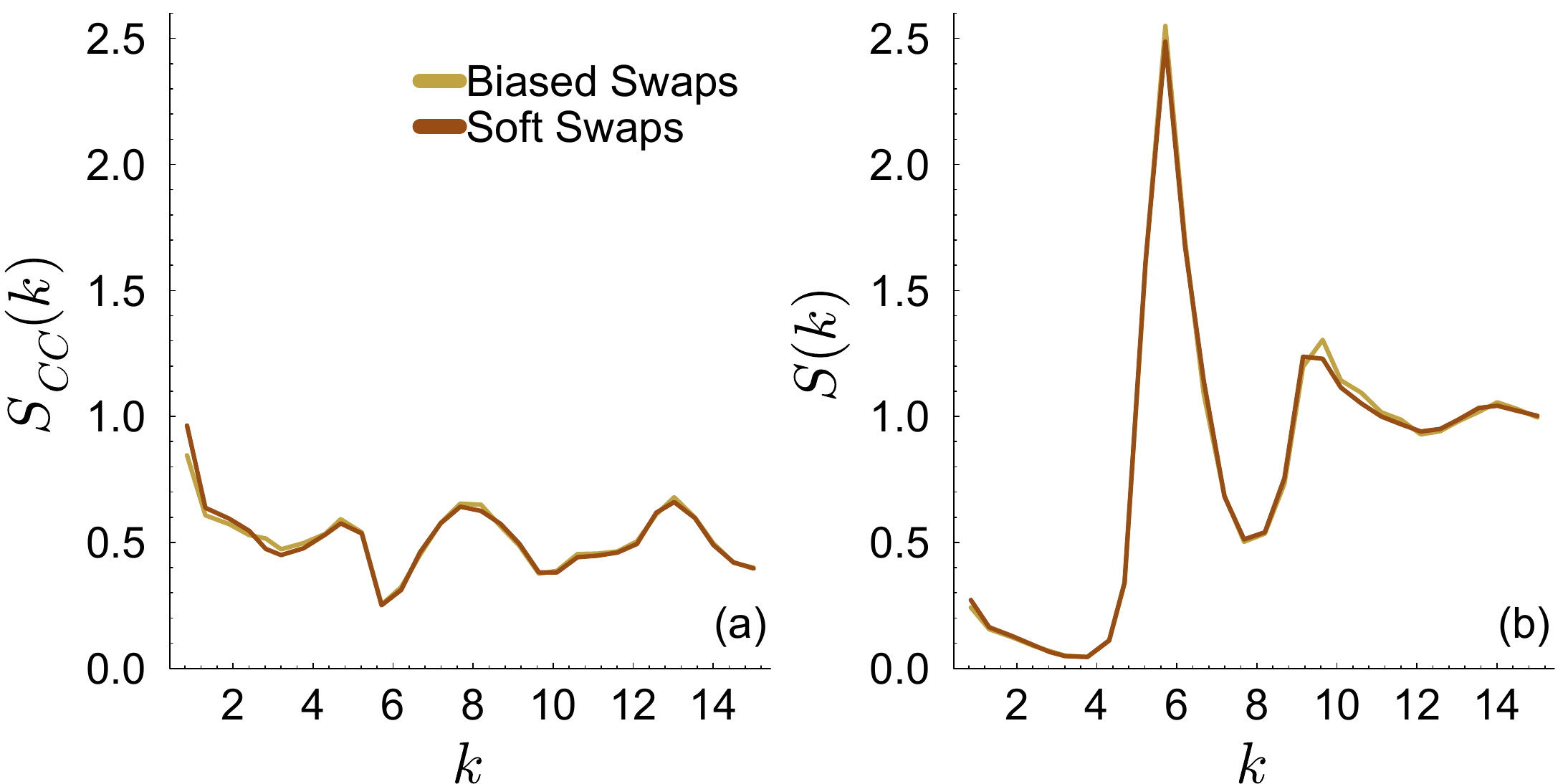}
    \caption{(a) Concentration-concentration structure factor $S_{CC}(k)$ and (b) total structure factor $S(k)$ at $T=0.155$ in the soft sphere model.}
    \label{fig:structure_factor}
\end{figure}

Finally, we also made sure that the systems did not crystallize or phase separate at low temperatures, by inspecting the relevant structure factors.
In particular, we compute the total structure factor $S(k)$ and the average concentration-concentration structure factor~\cite{Bhatia_Thornton_1970} $S_{CC}(k)=x_A^2 S_{AA}(k) + x_A^2 S_{BB}(k) - 2x_Bx_A S_{AB}(k)$ for trajectories.
In Fig.~\ref{fig:structure_factor}, we show representative results for the soft sphere model at the lowest temperature $T=0.155$: there are no signs of crystallization or phase separation in the shape of  $S_{CC}(k)$ and $S(k)$.

\section{Energy cost of swap moves}
\label{appendix:energy}

To better understand the difference in swap acceptance rates between the soft spheres and Kob-Andersen models, we carried out a spatially resolved analysis of the energy cost associated with a swap move.
We consider the swap between a particle of species $A$ and a particle of species $B$.
To simplify the analysis, we consider only the \textit{average} energy change due to such a swap.
Before the swap, the corresponding average single-particle energies of the two particles are $u_A$ and $u_B$, respectively, with
\beq
u_\alpha = \sum_\beta 4\pi\rho x_\alpha x_\beta\int_0^\infty r^2g_{\alpha\beta}(r)u_{\alpha\beta}(r)\,\mathrm{d}r,
\eeq
where $g_{\alpha\beta}(r)$ is the partial pair correlation function for particles of species $\alpha$ and $\beta$.

The corresponding average single-particle energies after the swap are denoted by $u_A'$ and $u_B'$, and the average energy difference due to the swap is $\Delta u_\alpha = u'_\alpha-u_\alpha$, with $\alpha=A, B$.
Since the particle positions remain the same when attempting the swap, we can write the average energy change for the particle of species $\alpha$ as
\begin{multline}
\Delta u_\alpha=\sum_{\beta}4\pi\rho\int_0^\infty x_\alpha x_\beta g_{\alpha\beta}(r)\times\\
\times\big(u_{\alpha'\beta}(r)-u_{\alpha'\alpha}(r)\big)\,\mathrm{d} r,
\label{eqn:integrand}
\end{multline}
where $\alpha$ and $\alpha'$ are the species before and after the swap, respectively.
The total average energy cost due to a swap is then calculated as $\Delta u=\Delta u_A+\Delta u_B$.

\begin{figure}[!tbp]
  \includegraphics[width=0.48\textwidth]{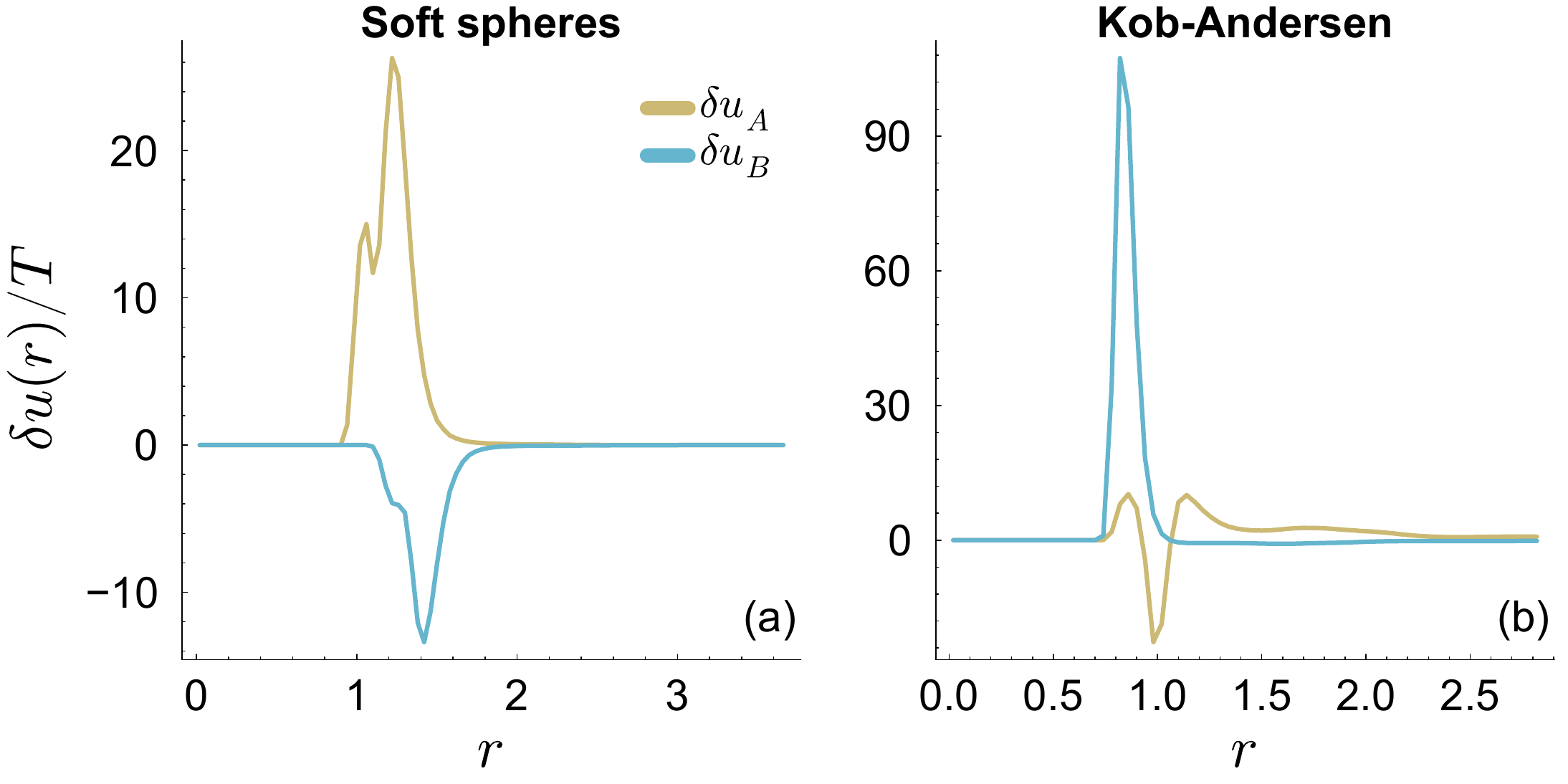}
  \caption{
    Integrands $\delta u_\alpha(r)$ of Eq.~\eqref{eqn:integrand} divided by $T$ for (a) the soft sphere model at $T=0.3$ and (b) the Kob-Andersen model at $T=0.8$.}
  \label{fig:swap_delta_energy}
\end{figure}
  
To get some insight into the contributions to $\Delta u$, we inspect the integrands $\delta u_\alpha\left(r\right)$ in Eq.~\eqref{eqn:integrand} as a function of $r$ using the actual $g_{\alpha\beta}(r)$ obtained from equilibrium configurations.
In Fig.~\ref{fig:swap_delta_energy}, we plot the two terms $\delta u_A$ and $\delta u_B$, normalized by the temperature, for our two benchmark models close to the onset temperature.
From the different signs of the integrands, we recognize that, as expected, what hinders particle swaps is the energy cost of expanding the small particle ($A$ for the soft spheres, $B$ for the Kob-Andersen model).
In the soft sphere model, the energy change due to shrinking a particle is always negative and has the largest contributions from particles at a distance of about 1.4, which corresponds to the most favorable distance between the $A$ and $B$ species.
By contrast, the energy change of the $A$-particles in the Kob-Andersen mixture has both positive and negative contributions, and the latter is negligible compared to the large energy cost of $\delta_B$.
The difference in acceptance rates between the two models likely reflects the discrepancies shown in Fig.~\ref{fig:swap_delta_energy}.
Our analysis also suggests that it may be possible to build a more efficient biased swap by targeting those small particles whose local environments have an excess of large particles in correspondence to the minima of $\delta_\alpha$, \textit{i.e.}, $r\approx 1.4$ for the soft sphere model and $r\approx 1$ for the Kob-Andersen mixture.
However, we expect the improvement to be negligible in this latter model due to the overall much lower acceptance rates.


%

\end{document}